\documentclass[aps, prl, reprint]{revtex4-2}

\usepackage{graphicx}

\usepackage[english]{babel}
\usepackage[utf8]{inputenc}
\usepackage{amsmath, amssymb, amsfonts}
\usepackage[none]{hyphenat}

\usepackage{hyperref}
\usepackage{color}

\begin{document}


    \title{Domino-cooling Oscillator Networks with Deep Reinforcement Learning}

    \author{Sampreet Kalita}
        \email{sampreet@alumni.iitg.ac.in}
    \author{Amarendra K. Sarma}
    \affiliation{Department of Physics, Indian Institute of Technology Guwahati, Guwahati-781039, Assam, India}

    \begin{abstract}
        The exploration of deep neural networks for optimal control has gathered a considerable amount of interest in recent years.
        Here, we utilize deep reinforcement learning to control individual evolutions of coupled harmonic oscillators in an oscillator network.
        Our work showcases a numerical approach to actively cool internal oscillators to their thermal ground states through modulated forces imparted to the external oscillators in the network.
        We present our results for thermal cooling of all oscillators in multiple network configurations and introduce the utility of our scheme in the quantum regime.
    \end{abstract}

    \maketitle

    \textit{Introduction.\textemdash{}}
        Controlling the effect of thermal noises in harmonic oscillator models often lies at the forefront of experiments that involve the study of quantum properties \cite{RevModPhys.82.1155}.
        Whereas typical experimental setups that investigate quantum phenomena use cryogenic temperatures in order to mitigate thermal effects of the external environment, numerous works have also reported the thermal cooling of oscillator models in the presence of environmental noises by employing the features of cavity optomechanics \cite{SBH.CavityOptomechanics.Aspelmeyer, NewJPhys.14.095015, NewJPhys.10.095009, PhysRevB.68.235328, PhysRevA.77.033804, PhysRevA.103.063509}.
        It may be noted here that most of the analytical methods which are used to obtain optimal cooling strategies utilize the dynamical evolution of ensemble-averaged variables, in which case, the stochastic behaviour of the inidividual trajectories and the noises associated with measurement are not fully captured.
        As such, active cooling techniques combining continuous external feedback with parametric modulations have been recently proposed to cool single trajectories of independent harmonic oscillators \cite{PhysRevA.107.053521, PhysRevA.107.023516}.
        However, for coupled harmonic oscillators, as the number of connections and their complexity increases, analytical derivations of the optimal expressions can become highly cumbersome.
        To circumvent this, numerical optimization using gradient-based techniques has recently emerged as a powerful tool \cite{PhysRevA.109.063508, PRXQuantum.4.030305}.
        In fact, it has been demonstrated that the cooling of motional modes in coupled oscillators with deterministic evolution can be hugely accelerated by implementing deep neural networks \cite{PhysRevResearch.4.L042038}.
        
        In the past decade, deep learning algorithms \cite{MIT.Goodfellow.DL} have been used to design and analyze several physical systems \cite{RevModPhys.91.045002, SciPostPhysLectNotes.29, NatRevPhys.5.141}.
        Examples include estimation of parameters and detecting phase transitions in condensed matter systems \cite{JPhys.33.053001, NatPhys.13.431}, predicting intermediate outputs of experimental runs and interpreting observations in high energy physics \cite{AnnuRevNuclPartSci.64.161, PhysLettB.778.64}, optimization of setups and image classification in optics \cite{NatCommun.9.4360, ApplPhysRev.7.021404}, and so on.
        They have also been employed to investigate quantum mechanical phenomena in qubits \cite{NewJPhys.22.045001, PRXQuantum.2.010316}, atomic ensembles \cite{PhysRevResearch.4.043216, Quantum.6.714, PhysRevApplied.14.014011} and hybrid quantum optical systems \cite{Optica.7.448, ApplPhysLett.118.164003}.
        The advantages of deep learning in the design of quantum circuits and the prediction of quantum measurements have also been reported \cite{QuantumSciTechnol.4.024004, PhysRevLett.127.140502, QuantumSciTechnol.8.025022}.
        It may be noted here that most of these applications utilize the paradigm of supervised learning \cite{Springer.Bishop.PRML, PhysRep.810.1}, where labelled data is used to train a neural network, typically for some prediction, or for the classification of unlabelled data.
        The paradigm of reinforcement learning (RL) \cite{MIT.Sutton.RL}, where a network learns to predict actions for each given set of observations, has only garnered interest in natural sciences in recent years, owing to its benefits in real-time, \textit{in-situ} workflows \cite{PhysRevA.107.010101, NIPS.2020.Interferobot}.
        Here, a software (or a harware) agent learns the optimal policy (strategy) of actions to direct particular properties of an environment (model) towards a goal.
        At each step, the agent receives a reward based on its actions.
        The incentive of higher cummulative rewards at the end of an episode results in the updation of its policy.
        Eventually, the agent starts choosing optimal actions from a finite action space at every step.
        For most physical systems, such an action space may represent the experimentally-realizable values of tunable parameters, such as the input power of laser drives or the duration of a pulse in optical setups, the alignment of logic gates in a quantum circuit, or the strength of external forces on a mechanical membrane.
        Accordingly, RL has applications in time-dependent quantum control \cite{PhysRevA.86.022321}, the prediction of optimal designs for quantum devices and experiments \cite{QuantumMachIntell.1.5, PhysRevLett.125.170501}, generation of non-classical states \cite{QuantumMachIntell.2.1, NatMachIntell.5.780}, and feedback assisted cooling of quantum systems \cite{PhysRevX.8.031084, SciRep.10.2623}.
        
        In this Letter, we present a deep RL-based approach to strategically cool a linearly-coupled network of harmonic oscillators via modulated external feedback forces.
        Our work improves the state-of-the-art schemes by three folds.
        First of all, it extends the prevalent cooling strategies for independent harmonic oscillators to a connected network of oscillators, where each network of fixed size can acquire different configurations.
        Secondly, our scheme cools down the internal oscillators of the network just by controlling the modulated feedback on the external ones, and retains system stability even in the presence of strong feedback.
        Finally, it can be used to constrain the eventual degree of cooling for each oscillator.
        Here, we demonstrate this RL-based approach for multiple network configurations and present its advantage in the deep quantum regime, by utilizing actor-critic approaches for deep learning \cite{ICML.2018.SAC, ICML.2020.TQC, ICLR.2022.DroQ, ICLR.2024.CrossQ}.

        \begin{figure}[!ht]
            \centering
            \includegraphics[width=0.48\textwidth]{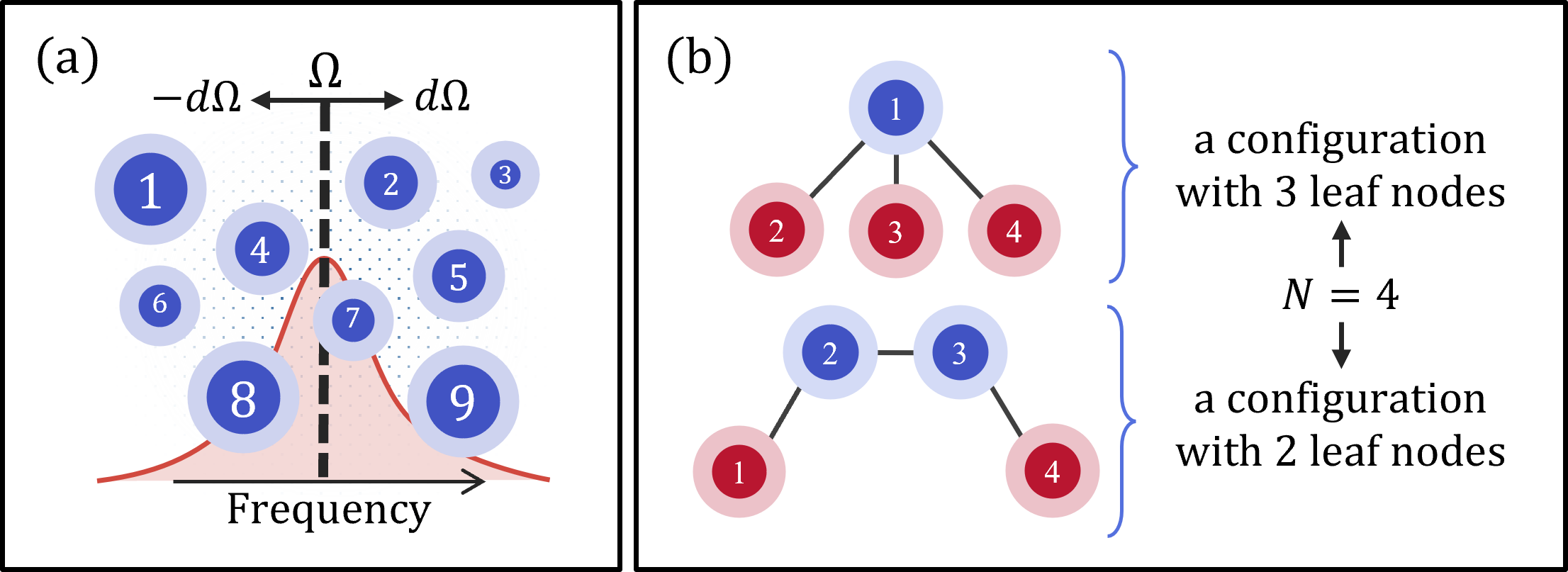}
            \caption{
                \textit{Network of harmonic oscillators:}
                (a) A collection of $N$ oscillators whose resonance frequences lie in the vicinity of $\Omega$.
                (b) Possible configurations with $4$ nodes and $3$ branches.
                Here, the leaf (internal) nodes are denoted in red (blue) color.
            }
            \label{fig:1}
        \end{figure}

    \textit{Setup.\textemdash{}}
        We consider a collection of $N$ nearly-identical harmonic oscillators (Fig. \ref{fig:1}(a)) coupled to one another \textit{either directly or indirectly}.
        The resultant network is a tree with $N$ nodes and $N - 1$ branches ($\mathcal{E}_{k} \equiv \{ (j_{k}, j_{k}^{\prime}) \}$, where $j \in [ 1, N ]$ denotes the oscillators' indices and $k \in [1, N - 1]$ the connections', unless specified otherwise).
        Such tree configurations contain one or more leaf nodes (a node connected to a maximum of one node) and may contain one or more internal nodes, thereby introducing a depth inside the oscillator network.
        For higher values of $N$, such networks can attain different configurations for the same $N$ (see Fig. \ref{fig:1}(b) for $N = 4$).
        We express all such configurations with the system Hamiltonian
        \begin{eqnarray}
            \label{eqn:hamiltonian}
            \frac{H}{\hbar} = \sum_{j} \Omega_{j} b_{j}^{\dagger} b_{j} - \sum_{k} \lambda \left( b_{j_{k}}^{\dagger} + b_{j_{k}} \right) \left( b_{j_{k}^{\prime}}^{\dagger} + b_{j_{k}^{\prime}} \right),
        \end{eqnarray}
        where $\Omega_{j}$ represent the frequencies of each constituent oscillator, that are centered around a base frequency $\Omega$ with deviations $\omega_{j}$ (sampled from a normal distribution of zero mean and standard deviation $d \Omega \ll \Omega$), $b_{j}$ and $b_{j}^{\dagger}$ are the Bosonic ladder operators, and $\lambda$ represents the strength of coupling between the connected oscillators.
        We now impart modulated kicks \textit{only} to the leaf nodes ($j^{*}$) of each configuration.
        The energy imparted by these kicks are depicted by the Hamiltonians $H_{j^{*}} = - \hbar \eta_{j^{*}} q_{j^{*}}^{2}$, with strengths $\eta_{j^{*}} = \eta \cos{( 2 \Omega t^{\prime} + \phi_{j^{*}} )}$, where $\phi_{j^{*}}$ are the feedback-controlled phases and $t^{\prime}$ is the origin-shifted time for each kicked interval.
        It might be important to note here that an analytical expression for optimal phase kicks to cool down independent oscillators (for $\lambda = 0$) is given by \cite{PhysRevA.107.053521} $\phi_{j^{*}}^{*} = \pi / 2 + \tanh{[ 1 / \{ p_{j^{*}} / q_{j^{*}} + \eta / ( 2 \Omega_{j^{*}} ) \} ]}$.
        Here, we train our RL-agent to predict the optimal phase kicks on the leaf nodes that can cool down the oscillators at \textit{any} node to \textit{target} thermal energies.

        \begin{figure}[!ht]
            \centering
            \includegraphics[width=0.48\textwidth]{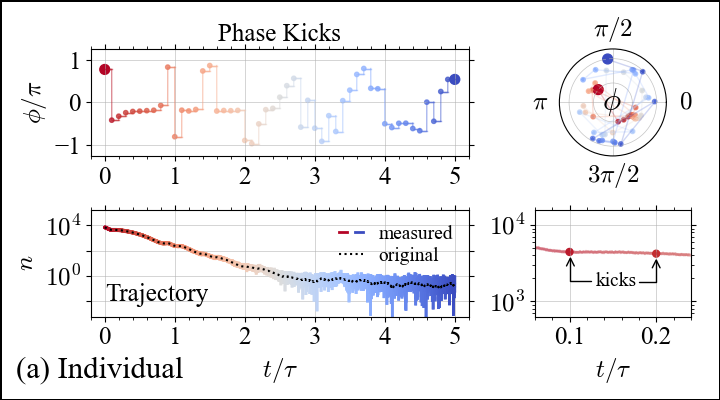}
            \includegraphics[width=0.48\textwidth]{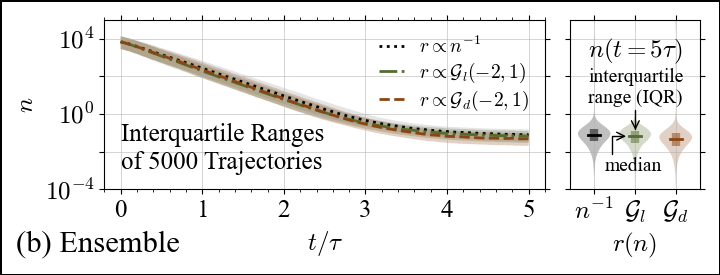}
            \caption{
                \textit{Cooling uncoupled oscillators:}
                (a) (clockwise from top left) Sequence of predicted phase kicks, $\phi$, in the presence of high measurement noises ($\sigma_{m} = 0.3$), it's representation in the polar form, illustration of the effect of the phase kicks and, dimensionless energies, $n = ( q^{2} + p^{2} ) / 2$, obtained after training for $10^{3}$ episodes.
                The RL-agent is rewarded through a difference function, $r \propto \mathcal{G}_{d} (\mu, \sigma) = \mathcal{G}_{l}^{(i)} (\mu, \sigma) - \mathcal{G}_{l}^{(i - 1)} (\mu, \sigma)$, such that $\mathcal{G}_{l}^{(i)} (\mu, \sigma) = \exp[\log_{10} [ n^{(i)} ] - \mu ] / ( 2 \sigma^{2} )$ is a Gaussian function evaluated at the $i$-th step of each episode for $\mu = - 2$ and $\sigma = 1$.
                Here, the dotted black line represents the original dynamics of the oscillator.
                (b) Interquartile ranges (IQRs) of $n$ (left panel) obtained after training and corresponding IQRs of $n$ at $t = 5 \tau$ (right) for different types of reward functions: inverse-energy ($r \propto n^{-1}$, dotted black line), Gaussian ($r \propto \mathcal{G}_{l} (\mu, \sigma)$, dash-dotted green line), difference ($r \propto \mathcal{G}_{d}$, dashed brown line).
                The IQRs are calculated with $10^{3}$ episodes after training for $10^{3}$ episodes with $5$ different seeds.
                Here, the base parameters are $\eta = \Omega / 2$, $\gamma = 10^{-6} \Omega$ and $\sigma_{m} = 0.1$.
                Initial values of position and momentum are sampled from a thermal distribution of $n_{\mathrm{th}} = 10^{4}$ such that $\langle q^{2} \rangle = \langle p^{2} \rangle = n_{\mathrm{th}} + 1 / 2$.
                Each trajectory starts with a unique frequency close to $\Omega$.
            }
            \label{fig:2}
        \end{figure}

    \textit{Reducing the thermal noise.\textemdash{}}
        For weakly dissipative oscillators ($\gamma \ll \Omega$, $\gamma$ denoting the damping rate of each oscillator), the dynamics of our model closely resemble the semi-classical dynamics until quantum fluctuations start dominating \cite{TaylorFrancis.QuantumOptomechanics.Bowen}.
        We therefore express the evolution of each oscillator as the stochastic differntial equations in terms of their dimensionless position and momentum quadratures ($q_{j} = ( b_{j}^{\dagger} + b_{j} ) / \sqrt{2}$ and $p_{j} = i ( b_{j}^{\dagger} - b_{j} ) / \sqrt{2}$):
        \begin{subequations}
            \label{eqn:evolution}
            \begin{eqnarray}
                \dot{q}_{j} & = & - \frac{\gamma}{2} q_{j} + \Omega_{j} p_{j} + \sqrt{\gamma} q_{j}^{(in)}, \\
                \dot{p}_{j} & = &  - \frac{\gamma}{2} p_{j} - \Omega_{j} q_{j} + \sum_{k^{*}} \lambda q_{j_{k^{*}}^{\prime}} + 2 \eta_{j^{*}} q_{j^{*}} + \sqrt{\gamma} p_{j}^{(in)},
            \end{eqnarray}
        \end{subequations}
        where $k^{*}$ represent those branches that have $q_{j}$ as an end-node, and $j^{*}$ represent the leaf nodes.
        Here, $q_{j}^{(in)}$ and $p_{j}^{(in)}$ are the canonical noises entering each mode via the thermal environment \cite{CUP.QuantumOptics.Agarwal}.
        We can further write Eqs. \eqref{eqn:evolution} in a compact form: $dv(t) = f(t, v(t)) dt + g dw(t)$, where $v = (q_{1}, p_{1}, q_{2}, p_{2}, \dots, q_{N}, p_{N})^{T}$ is the vector of the quadratures, $f$ is the $2 N \times 2 N$ drift matrix, $g = \textrm{Diag} [ \sqrt{\gamma (n_{\mathrm{th}} + 1 / 2)}, \sqrt{\gamma (n_{\mathrm{th}} + 1 / 2)}, \dots ]$ is the diffusion matrix and $dw$ are the $2 N$ noise control terms.

        \begin{figure*}[!ht]
            \centering
            \includegraphics[width=0.3\textwidth]{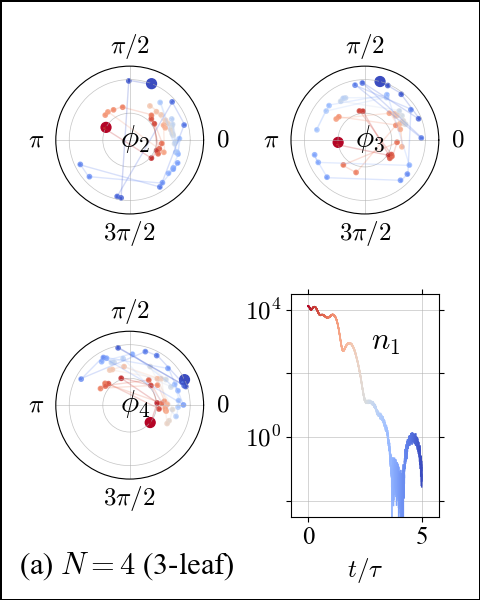}
            \includegraphics[width=0.3\textwidth]{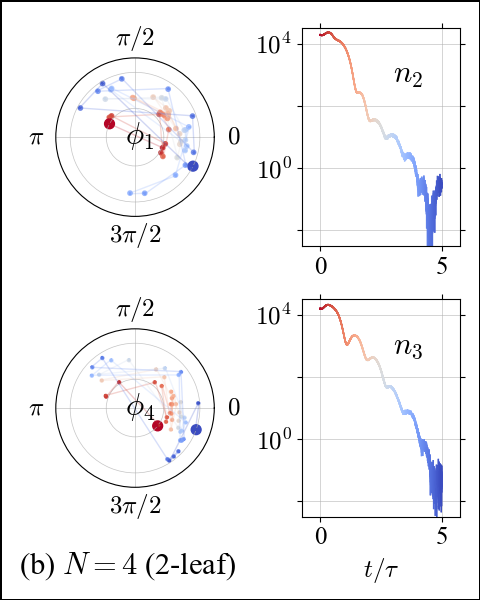}
            \includegraphics[width=0.3\textwidth]{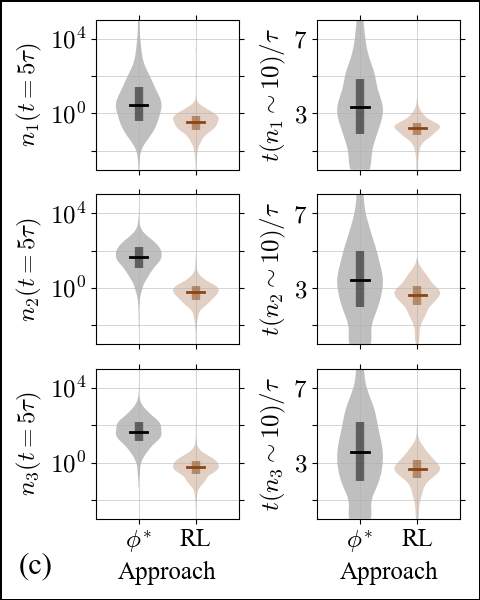}
            \caption{
                \textit{Cooling coupled oscillators:}
                Feedback-phase kicks, $\phi_{j}$, imparted on the leaf nodes and time evolution of dimensionless energies, $n_{j}$, of the internal nodes for $N = 4$ with (a) a 3-leaf configuration and (b) a 2-leaf configuration, obtained after the learning curves saturate.
                (c) Comparison of analytical ($\phi^{*}$, black) and RL-based (RL, brown) approaches: IQRs of final energies at $t = 5 \tau$ ($n_{j}$, left panels) and time required to reach $n_{j} \sim 10$ ($t$, right panels).
                Here, $\lambda = \Omega / 2$, $r \propto \mathcal{G}_{d} (-1, 2)$ and each episode is initialized with a unique set of $\Omega_{j}$.
                The IQRs are calculated with $10^{3}$ episodes.
                Other parameters are the same as Fig. \ref{fig:2}(b).
            }
            \label{fig:3}
        \end{figure*}

        For our semi-classical analysis, we choose the noises $dw$ from a normal distribution with zero mean and variance $\sqrt{dt}$ such that the evolution of $v$ follows a Wiener process.
        Further, to mimic measurement imprecision, we overlay a Gaussian noise of zero mean and variance $\sigma_{m}$ on the original dynamics obtained from Eqs. \eqref{eqn:evolution}.
        At an interval of $0.1 \tau$ ($\tau = 2 \pi / \Omega$), these \textit{measured} values of positions, $q_{j}$, and momentums, $p_{j}$, are utilized to calculate a reward metric.
        We design this metric in such a way that heating is penalized and cooling is rewarded.
        The RL-agent thus learns to maximize the cumulative reward obtained at the end of each episode (trajectory).
        Intricate details of this implementation are mentioned in Ref. \cite{suppl}.

        We briefly demonstrate this approach for independent oscillator models.
        As depicted in Fig. \ref{fig:2}(a), the RL-agent successfully learns an optimal policy for the phase kicks to successively cool the oscillators below a thermal energy of $\hbar \Omega$, even in the presence of high measurement noises.
        Beyond this, the stochastic noise effects start dominating and the subsequent regime depicts the noisy equilibrium dynamics.
        It may be noted here that the rate of cooling in this scenario closely matches with that of the analytically obtained optimal sequences proposed in the Refs. \cite{PhysRevA.107.053521, PhysRevA.107.023516}.
        A slight improvement in our scheme can be attributed to the inclusion of measurement imprecisions which are not taken into account while deriving the optimal analytical expressions for feedback-cooling.
        In addition to this, one can control the degree of cooling for each oscillator upto a fixed value of energy by using a bounded reward function with a certain degree of relaxation, as shown in Fig. \ref{fig:2}(b).
        Moreover, as this scheme is generalized for all mechanical frequencies in the vicinity of the base frequency $\Omega$, it can also be used to simultaneously cool down a collection of nearly-identical independent oscillators.

        We now discuss the scenario for the oscillator network.
        Here, one can have multiple configurations, such as the 3-leaf or 2-leaf configurations for $N = 4$ (see Fig. \ref{fig:1}(b)).
        We illustrate the results for the 3-leaf case in Fig. \ref{fig:3}(a).
        Through a non-intuitive sequence of phase kicks to the two oscillators on the leaf nodes, the RL-agent learns to cool down the internal oscillators below unit normalized energies.
        It is worth mentioning here that the modulated forces are only applied to the leaf nodes and the internal nodes are indirectly cooled through the application of the adaptive phase kicks in these forces.
        A similar pattern is observed for the 2-leaf case too, as depicted in Fig. \ref{fig:3}(b).
        The RL-agent learns to cool down the undriven internal oscillator through a unique series of phase kicks imparted to the two leaf oscillators.
        In Fig. \ref{fig:3}(c), we compare the effectiveness of the RL-based approach with that of the analytical approach of Ref. \cite{PhysRevA.107.053521}.
        As seen from the figure, the rate of cooling is higher in the RL-based approach.
        It may be noted here that even with a strong feedback force on the leaf nodes, the system avoids parametric instability as an additional outcome of the cooling policy.
        Furthermore, the feedback forces can also be updated at irregular intervals as the RL-agent eventually learns to predict the optimal actions for \textit{any} set of observations.
        This illustrates the robustness of the RL-based approach.

        \begin{figure*}[!ht]
            \centering
            \includegraphics[width=0.94\textwidth]{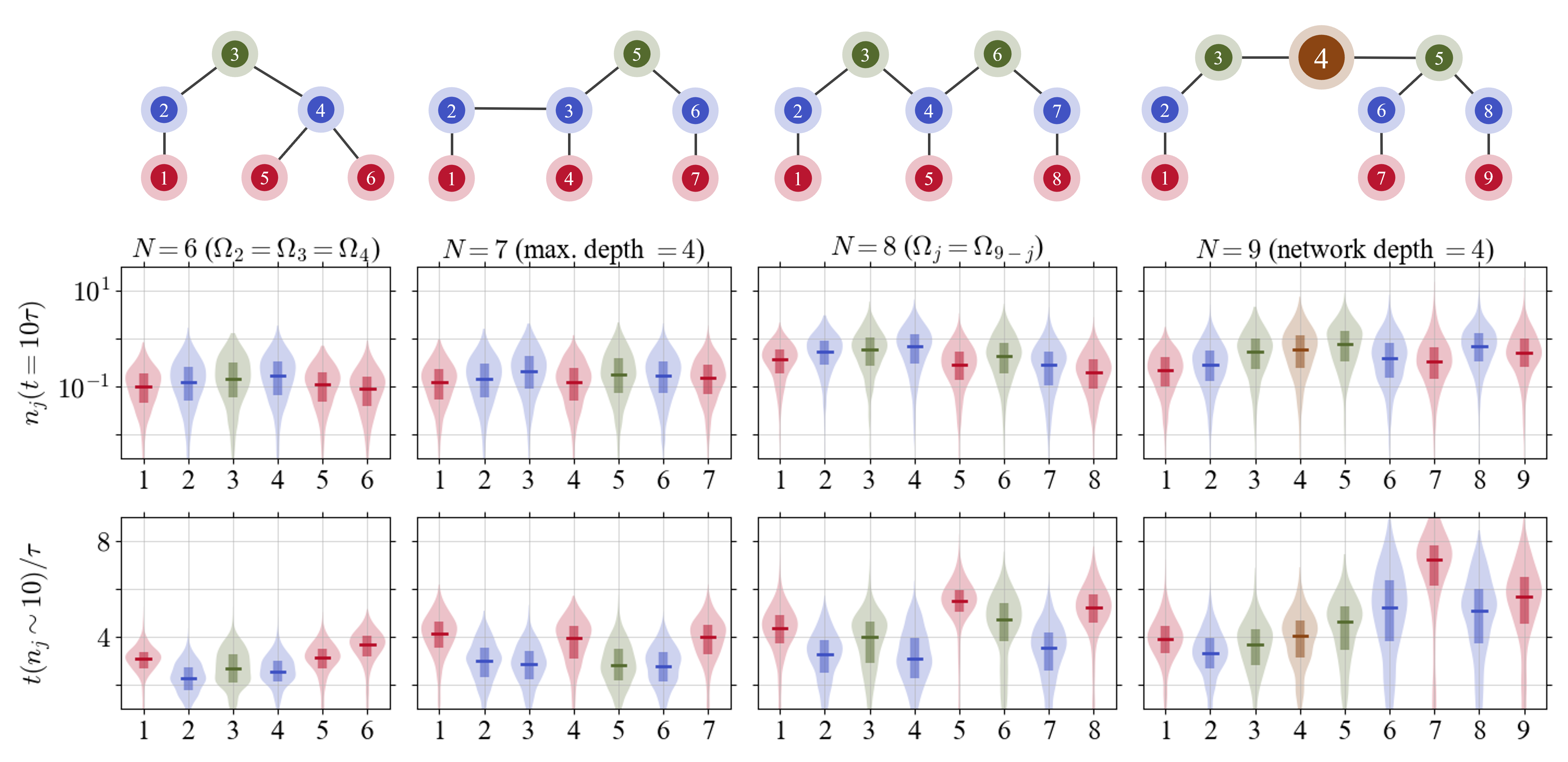}
            \caption{
                \textit{Cooling special configurations:} IQRs of energies at $t = 10 \tau$ ($n_j$, center panels) and time required to reach $n_{j} = 10$ ($t$, bottom panels) for different configurations with higher values of $N$ (top panels) obtained after the learning curves saturate.
                The red colored nodes represent the leaf oscillators, the ones in blue represent the internal oscillators connected to at least one leaf node, and the rest represent the oscillators with depth $> 2$.
                Here, $\Omega_{j} / \Omega \approx [ 1.119, 0.974, 0.974, 0.974, 0.983, 1.016 ]$ for $N = 6$, $[ 1.192, 0.944, 1.096, 0.943, 0.992, 1.015, 1.238 ]$ for $N = 7$, $[ 1.043, 0.879, 1.056, 1.019, 1.019, 1.056, 0.879, 1.043 ]$ for $N = 8$ and $[ 0.920, 0.796, 1.060, 1.074, 0.969, 1.036, 1.171, 1.106, 1.070 ]$ for $N = 9$.
                Other parameters are the same as Fig. \ref{fig:3}.
            }
            \label{fig:4}
        \end{figure*}

        For typical situations, where the oscillators are formed using mechanical degrees of freedom, the feedback forces may be thought of as spring-constant modulations \cite{NatCommun.11.1589}.
        The leaf-node oscillators may also depict detuned optical (microwave) modes which are further coupled to internal nodes having mechanical degrees of freedom with lower decay rates \cite{SBH.CavityOptomechanics.Aspelmeyer}.
        The external forces in such models are analogous to squeezed optical (microwave) drives \cite{PhysRevA.89.023849, PhysRevLett.107.043603, PhysRevLett.111.073603}.

        Nonetheless, the RL-based approach presented in this letter provides a systematic way to cool down the internal oscillators in a deep network of linearly coupled harmonic oscillators by strategically manipulating the dynamics of the external ones through the  phase-adaptive feedback.
        To demonstrate the scalability of this approach, we now extend our formalism to higher values of $N$ and plot the distribution of the final energies at $t = 10 \tau$ for different types of configurations in Fig. \ref{fig:4}.
        Here, we observe that, given a suitable stretch of training, the RL-agent learns to cool down the internal nodes in these configurations, with an approximate duration to reach the dimensionless energy of $10$ being under $5 \tau$.
        It may be noted here that beyond this value, the quantum jumps play an important role and a semi-classical approach may not necessarily describe the actual quantum behaviour of the network.
        Here, we briefly present the applicability of our RL-based approach in steering the dynamics of quantum harmonic oscillators by implementing a fully quantum treatment.

\begin{figure}[!h]
            \centering
            \includegraphics[width=0.48\textwidth]{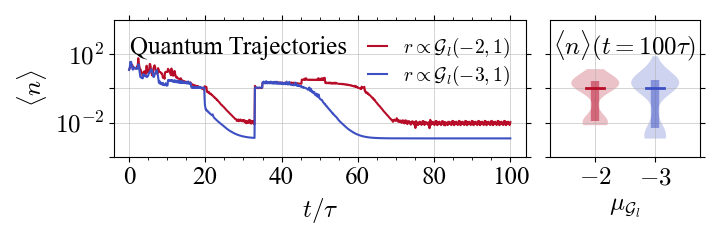}
            \caption{
                \textit{Towards quantum control:}
                Bosonic occupancies for independent harmonic oscillators in the presence of quantum fluctuations for two different target occupancies ($10^{\mu}$) and their corresponding IQRs at $t = 100 \tau$.
                The reward function used is $r \propto \mathcal{G}_{l} ( \mu, \sigma )$ with $\mu = -2$ (red) and $\mu = -3$ (blue).
                Here, the initial state is a Fock state with $n_{0} = 12$. 
                Other parameters are the same as Figs. \ref{fig:2}(b).
            }
            \label{fig:5}
        \end{figure}

    \textit{Controlling quantum fluctuations.\textemdash{}}
        We now consider the stochastic evolution of the oscillators' quantum state using Monte-Carlo quantum trajectories \cite{AmJPhys.70.719}, where each trajectory represents a single experimental run.
        For this case, the effective non-Hermitian Hamiltonian is given by
        \begin{eqnarray}
            \label{eqn:hamiltonian_effective}
            H_{\mathrm{eff}} = H + \sum_{j^{*}} H_{j^{*}} - \frac{i \hbar}{2} \sum_{j, \pm} C_{j}^{(\pm) \dagger} C_{j}^{(\pm)},
        \end{eqnarray}
        where $C_{j}^{(+)} = \sqrt{\gamma n_{\mathrm{th}}} b_{j}^{\dagger}$ and $C_{j}^{(-)} = \sqrt{\gamma ( n_{\mathrm{th}} + 1 ) } b_{j}$ are the quantum jump operators.
        The occurence of a jump, and the operator that causes it, are stochastically chosen by using random variables from a uniform distribution.
        If the state does not collapse, its evolution is described by the Hamiltonian in Eq. \eqref{eqn:hamiltonian_effective} until the next jump \cite{ComputPhysCommun.183.1760, ComputPhysCommun.184.1234}.
        For this scenario, the RL-agent's observations comprise of $\langle q_{j}^{2} \rangle$, $\langle p_{j}^{2} \rangle$ and $\langle q_{j} p_{j} \rangle$, which are obtained from the state of the oscillator and its reward is calculated using $\langle n_{j} \rangle = ( \langle q_{j}^{2} \rangle + \langle p_{j}^{2} \rangle - 1 ) / 2$ in a similar way as described for the semi-classical case.
        In Fig. \ref{fig:5}, we demonstrate the control over the dynamics of single quantum trajectories by using two different reward functions.
        It can be seen here that even in the presence of quantum fluctuations, the phase kicks predicted by the RL-based approach for individual trajetories can cool the oscillators to target occupancies beyond the feedback cooling limit of optimal control techniques for ensembled averages \cite{PhysRevA.107.053521, PhysRevA.109.063508}.

        In what follows, we mention a few challenges faced in our RL-based imlementation and the future direction of our research.
        The primary bottleneck of this approach is the training time.
        Although soft-actor critic algorithms are guaranteed to converge \cite{ICML.2018.SAC}, the time needed to reach this convergence may range from a few minutes (for the individual oscillators) to a handful of hours (for $N \geq 4$).
        In addition to this, the degree of precision demanded by the Wiener process and the probabilistic workflow of the quantum Monte-Carlo method extends the computation time for each steps substantially.
        It may be noted here that our implementations are GPU-accelerated and we utilize the JAX ecosystem with our library \texttt{quantrl} \cite{quantrl}.
        Secondly, although we have selected reasonable numerical values for the action step sizes, in typical experimental setups, the hardware that executes the predicted actions may introduce additional latencies resulting in delayed actions.
        In future works, we intend to incorporate such delayed or asynchronous actions and include the effect of measurement-induced collapes in quantum trajectories.

    \textit{Conclusions.\textemdash{}}
        We proposed a generalized scheme for the simultaneous cooling of coupled harmonic oscillators in an oscillator network, by using periodically-modulated external forces whose feedback-phases were predicted by a deep reinforcement learning-based numerical model.
        Notably, we imparted these external forces only to the leaf nodes of the network and still achieved a significant amount of cooling of all its nodes.
        The degree of cooling is comparable to the cooling achieved by state-of-the-art analytical techniques for uncoupled oscillators.
        Further, we discussed the robustness of our approach to the depth of the network and its validity even in the deep quantum regime.
        Our work may find applications in experimental workflows where thermal cooling is necessary to devise quantum systems, or for studying quantum phenomena.

    S.K. conceptualized the framework, implemented the approaches and performed the numerical experiments.
    A.K.S. scrutinized the writing and discussed the results.

    S.K. acknowledges S. Chakraborty and P. Kumar for insightful discussions and comments regarding the work.
    S.K. acknowledges MHRD, Government of India and Google LLC for supporting the work through the PMRF scheme and GCP credits respectively.
    A.K.S. acknowledges the grant from MoE, Government of India (Grant no. MoE-STARS/STARS-2/2023-0161).

    All the libraries related to the work are open-sourced on GitHub \cite{quantrl}.
    The scripts for the oscillator networks and corresponding data for the plots are available upon reasonable request to the authors.

    \bibliography{references}

\begin{thebibliography}{59}%
\makeatletter
\providecommand \@ifxundefined [1]{%
 \@ifx{#1\undefined}
}%
\providecommand \@ifnum [1]{%
 \ifnum #1\expandafter \@firstoftwo
 \else \expandafter \@secondoftwo
 \fi
}%
\providecommand \@ifx [1]{%
 \ifx #1\expandafter \@firstoftwo
 \else \expandafter \@secondoftwo
 \fi
}%
\providecommand \natexlab [1]{#1}%
\providecommand \enquote  [1]{``#1''}%
\providecommand \bibnamefont  [1]{#1}%
\providecommand \bibfnamefont [1]{#1}%
\providecommand \citenamefont [1]{#1}%
\providecommand \href@noop [0]{\@secondoftwo}%
\providecommand \href [0]{\begingroup \@sanitize@url \@href}%
\providecommand \@href[1]{\@@startlink{#1}\@@href}%
\providecommand \@@href[1]{\endgroup#1\@@endlink}%
\providecommand \@sanitize@url [0]{\catcode `\\12\catcode `\$12\catcode
  `\&12\catcode `\#12\catcode `\^12\catcode `\_12\catcode `\%12\relax}%
\providecommand \@@startlink[1]{}%
\providecommand \@@endlink[0]{}%
\providecommand \url  [0]{\begingroup\@sanitize@url \@url }%
\providecommand \@url [1]{\endgroup\@href {#1}{\urlprefix }}%
\providecommand \urlprefix  [0]{URL }%
\providecommand \Eprint [0]{\href }%
\providecommand \doibase [0]{https://doi.org/}%
\providecommand \selectlanguage [0]{\@gobble}%
\providecommand \bibinfo  [0]{\@secondoftwo}%
\providecommand \bibfield  [0]{\@secondoftwo}%
\providecommand \translation [1]{[#1]}%
\providecommand \BibitemOpen [0]{}%
\providecommand \bibitemStop [0]{}%
\providecommand \bibitemNoStop [0]{.\EOS\space}%
\providecommand \EOS [0]{\spacefactor3000\relax}%
\providecommand \BibitemShut  [1]{\csname bibitem#1\endcsname}%
\let\auto@bib@innerbib\@empty
\bibitem [{\citenamefont {Clerk}\ \emph {et~al.}(2010)\citenamefont {Clerk},
  \citenamefont {Devoret}, \citenamefont {Girvin}, \citenamefont {Marquardt},\
  and\ \citenamefont {Schoelkopf}}]{RevModPhys.82.1155}%
  \BibitemOpen
  \bibfield  {author} {\bibinfo {author} {\bibfnamefont {A.~A.}\ \bibnamefont
  {Clerk}}, \bibinfo {author} {\bibfnamefont {M.~H.}\ \bibnamefont {Devoret}},
  \bibinfo {author} {\bibfnamefont {S.~M.}\ \bibnamefont {Girvin}}, \bibinfo
  {author} {\bibfnamefont {F.}~\bibnamefont {Marquardt}},\ and\ \bibinfo
  {author} {\bibfnamefont {R.~J.}\ \bibnamefont {Schoelkopf}},\ }\href
  {https://doi.org/10.1103/RevModPhys.82.1155} {\bibfield  {journal} {\bibinfo
  {journal} {Rev. Mod. Phys.}\ }\textbf {\bibinfo {volume} {82}},\ \bibinfo
  {pages} {1155} (\bibinfo {year} {2010})}\BibitemShut {NoStop}%
\bibitem [{\citenamefont {Aspelmeyer}\ \emph {et~al.}(2014)\citenamefont
  {Aspelmeyer}, \citenamefont {Kippenberg},\ and\ \citenamefont
  {Marquardt}}]{SBH.CavityOptomechanics.Aspelmeyer}%
  \BibitemOpen
  \bibfield  {author} {\bibinfo {author} {\bibfnamefont {M.}~\bibnamefont
  {Aspelmeyer}}, \bibinfo {author} {\bibfnamefont {T.~J.}\ \bibnamefont
  {Kippenberg}},\ and\ \bibinfo {author} {\bibfnamefont {F.}~\bibnamefont
  {Marquardt}},\ }\href {https://books.google.co.in/books?id=FG71AwAAQBAJ}
  {\emph {\bibinfo {title} {Cavity Optomechanics: Nano- and Micromechanical
  Resonators Interacting with Light}}}\ (\bibinfo  {publisher} {Springer Berlin
  Heidelberg},\ \bibinfo {year} {2014})\BibitemShut {NoStop}%
\bibitem [{\citenamefont {Karuza}\ \emph {et~al.}(2012)\citenamefont {Karuza},
  \citenamefont {Molinelli}, \citenamefont {Galassi}, \citenamefont
  {Biancofiore}, \citenamefont {Natali}, \citenamefont {Tombesi}, \citenamefont
  {Di~Giuseppe},\ and\ \citenamefont {Vitali}}]{NewJPhys.14.095015}%
  \BibitemOpen
  \bibfield  {author} {\bibinfo {author} {\bibfnamefont {M.}~\bibnamefont
  {Karuza}}, \bibinfo {author} {\bibfnamefont {C.}~\bibnamefont {Molinelli}},
  \bibinfo {author} {\bibfnamefont {M.}~\bibnamefont {Galassi}}, \bibinfo
  {author} {\bibfnamefont {C.}~\bibnamefont {Biancofiore}}, \bibinfo {author}
  {\bibfnamefont {R.}~\bibnamefont {Natali}}, \bibinfo {author} {\bibfnamefont
  {P.}~\bibnamefont {Tombesi}}, \bibinfo {author} {\bibfnamefont
  {G.}~\bibnamefont {Di~Giuseppe}},\ and\ \bibinfo {author} {\bibfnamefont
  {D.}~\bibnamefont {Vitali}},\ }\href@noop {} {\bibfield  {journal} {\bibinfo
  {journal} {New Journal of Physics}\ }\textbf {\bibinfo {volume} {14}},\
  \bibinfo {pages} {095015} (\bibinfo {year} {2012})}\BibitemShut {NoStop}%
\bibitem [{\citenamefont {Genes}\ \emph
  {et~al.}(2008{\natexlab{a}})\citenamefont {Genes}, \citenamefont {Vitali},\
  and\ \citenamefont {Tombesi}}]{NewJPhys.10.095009}%
  \BibitemOpen
  \bibfield  {author} {\bibinfo {author} {\bibfnamefont {C.}~\bibnamefont
  {Genes}}, \bibinfo {author} {\bibfnamefont {D.}~\bibnamefont {Vitali}},\ and\
  \bibinfo {author} {\bibfnamefont {P.}~\bibnamefont {Tombesi}},\ }\href
  {https://doi.org/10.1088/1367-2630/10/9/095009} {\bibfield  {journal}
  {\bibinfo  {journal} {New Journal of Physics}\ }\textbf {\bibinfo {volume}
  {10}},\ \bibinfo {pages} {095009} (\bibinfo {year}
  {2008}{\natexlab{a}})}\BibitemShut {NoStop}%
\bibitem [{\citenamefont {Hopkins}\ \emph {et~al.}(2003)\citenamefont
  {Hopkins}, \citenamefont {Jacobs}, \citenamefont {Habib},\ and\ \citenamefont
  {Schwab}}]{PhysRevB.68.235328}%
  \BibitemOpen
  \bibfield  {author} {\bibinfo {author} {\bibfnamefont {A.}~\bibnamefont
  {Hopkins}}, \bibinfo {author} {\bibfnamefont {K.}~\bibnamefont {Jacobs}},
  \bibinfo {author} {\bibfnamefont {S.}~\bibnamefont {Habib}},\ and\ \bibinfo
  {author} {\bibfnamefont {K.}~\bibnamefont {Schwab}},\ }\href
  {https://doi.org/10.1103/PhysRevB.68.235328} {\bibfield  {journal} {\bibinfo
  {journal} {Phys. Rev. B}\ }\textbf {\bibinfo {volume} {68}},\ \bibinfo
  {pages} {235328} (\bibinfo {year} {2003})}\BibitemShut {NoStop}%
\bibitem [{\citenamefont {Genes}\ \emph
  {et~al.}(2008{\natexlab{b}})\citenamefont {Genes}, \citenamefont {Vitali},
  \citenamefont {Tombesi}, \citenamefont {Gigan},\ and\ \citenamefont
  {Aspelmeyer}}]{PhysRevA.77.033804}%
  \BibitemOpen
  \bibfield  {author} {\bibinfo {author} {\bibfnamefont {C.}~\bibnamefont
  {Genes}}, \bibinfo {author} {\bibfnamefont {D.}~\bibnamefont {Vitali}},
  \bibinfo {author} {\bibfnamefont {P.}~\bibnamefont {Tombesi}}, \bibinfo
  {author} {\bibfnamefont {S.}~\bibnamefont {Gigan}},\ and\ \bibinfo {author}
  {\bibfnamefont {M.}~\bibnamefont {Aspelmeyer}},\ }\href
  {https://doi.org/10.1103/PhysRevA.77.033804} {\bibfield  {journal} {\bibinfo
  {journal} {Phys. Rev. A}\ }\textbf {\bibinfo {volume} {77}},\ \bibinfo
  {pages} {033804} (\bibinfo {year} {2008}{\natexlab{b}})}\BibitemShut
  {NoStop}%
\bibitem [{\citenamefont {Lai}\ \emph {et~al.}(2021)\citenamefont {Lai},
  \citenamefont {Huang}, \citenamefont {Hou}, \citenamefont {Nori},\ and\
  \citenamefont {Liao}}]{PhysRevA.103.063509}%
  \BibitemOpen
  \bibfield  {author} {\bibinfo {author} {\bibfnamefont {D.-G.}\ \bibnamefont
  {Lai}}, \bibinfo {author} {\bibfnamefont {J.}~\bibnamefont {Huang}}, \bibinfo
  {author} {\bibfnamefont {B.-P.}\ \bibnamefont {Hou}}, \bibinfo {author}
  {\bibfnamefont {F.}~\bibnamefont {Nori}},\ and\ \bibinfo {author}
  {\bibfnamefont {J.-Q.}\ \bibnamefont {Liao}},\ }\href
  {https://doi.org/10.1103/PhysRevA.103.063509} {\bibfield  {journal} {\bibinfo
   {journal} {Phys. Rev. A}\ }\textbf {\bibinfo {volume} {103}},\ \bibinfo
  {pages} {063509} (\bibinfo {year} {2021})}\BibitemShut {NoStop}%
\bibitem [{\citenamefont {Ghosh}\ \emph {et~al.}(2023)\citenamefont {Ghosh},
  \citenamefont {Kumar}, \citenamefont {Sommer}, \citenamefont {Jimenez},
  \citenamefont {Sudhir},\ and\ \citenamefont {Genes}}]{PhysRevA.107.053521}%
  \BibitemOpen
  \bibfield  {author} {\bibinfo {author} {\bibfnamefont {A.}~\bibnamefont
  {Ghosh}}, \bibinfo {author} {\bibfnamefont {P.}~\bibnamefont {Kumar}},
  \bibinfo {author} {\bibfnamefont {C.}~\bibnamefont {Sommer}}, \bibinfo
  {author} {\bibfnamefont {F.~G.}\ \bibnamefont {Jimenez}}, \bibinfo {author}
  {\bibfnamefont {V.}~\bibnamefont {Sudhir}},\ and\ \bibinfo {author}
  {\bibfnamefont {C.}~\bibnamefont {Genes}},\ }\href
  {https://doi.org/10.1103/PhysRevA.107.053521} {\bibfield  {journal} {\bibinfo
   {journal} {Phys. Rev. A}\ }\textbf {\bibinfo {volume} {107}},\ \bibinfo
  {pages} {053521} (\bibinfo {year} {2023})}\BibitemShut {NoStop}%
\bibitem [{\citenamefont {Manikandan}\ and\ \citenamefont
  {Qvarfort}(2023)}]{PhysRevA.107.023516}%
  \BibitemOpen
  \bibfield  {author} {\bibinfo {author} {\bibfnamefont {S.~K.}\ \bibnamefont
  {Manikandan}}\ and\ \bibinfo {author} {\bibfnamefont {S.}~\bibnamefont
  {Qvarfort}},\ }\href {https://doi.org/10.1103/PhysRevA.107.023516} {\bibfield
   {journal} {\bibinfo  {journal} {Phys. Rev. A}\ }\textbf {\bibinfo {volume}
  {107}},\ \bibinfo {pages} {023516} (\bibinfo {year} {2023})}\BibitemShut
  {NoStop}%
\bibitem [{\citenamefont {Liu}\ \emph {et~al.}(2024)\citenamefont {Liu},
  \citenamefont {Zeng}, \citenamefont {Tan}, \citenamefont {Dong},
  \citenamefont {Nori},\ and\ \citenamefont {Liao}}]{PhysRevA.109.063508}%
  \BibitemOpen
  \bibfield  {author} {\bibinfo {author} {\bibfnamefont {Y.-H.}\ \bibnamefont
  {Liu}}, \bibinfo {author} {\bibfnamefont {Y.}~\bibnamefont {Zeng}}, \bibinfo
  {author} {\bibfnamefont {Q.-S.}\ \bibnamefont {Tan}}, \bibinfo {author}
  {\bibfnamefont {D.}~\bibnamefont {Dong}}, \bibinfo {author} {\bibfnamefont
  {F.}~\bibnamefont {Nori}},\ and\ \bibinfo {author} {\bibfnamefont {J.-Q.}\
  \bibnamefont {Liao}},\ }\href {https://doi.org/10.1103/PhysRevA.109.063508}
  {\bibfield  {journal} {\bibinfo  {journal} {Phys. Rev. A}\ }\textbf {\bibinfo
  {volume} {109}},\ \bibinfo {pages} {063508} (\bibinfo {year}
  {2024})}\BibitemShut {NoStop}%
\bibitem [{\citenamefont {Porotti}\ \emph {et~al.}(2023)\citenamefont
  {Porotti}, \citenamefont {Peano},\ and\ \citenamefont
  {Marquardt}}]{PRXQuantum.4.030305}%
  \BibitemOpen
  \bibfield  {author} {\bibinfo {author} {\bibfnamefont {R.}~\bibnamefont
  {Porotti}}, \bibinfo {author} {\bibfnamefont {V.}~\bibnamefont {Peano}},\
  and\ \bibinfo {author} {\bibfnamefont {F.}~\bibnamefont {Marquardt}},\ }\href
  {https://doi.org/10.1103/PRXQuantum.4.030305} {\bibfield  {journal} {\bibinfo
   {journal} {PRX Quantum}\ }\textbf {\bibinfo {volume} {4}},\ \bibinfo {pages}
  {030305} (\bibinfo {year} {2023})}\BibitemShut {NoStop}%
\bibitem [{\citenamefont {Sarma}\ \emph {et~al.}(2022)\citenamefont {Sarma},
  \citenamefont {Borah}, \citenamefont {Kani},\ and\ \citenamefont
  {Twamley}}]{PhysRevResearch.4.L042038}%
  \BibitemOpen
  \bibfield  {author} {\bibinfo {author} {\bibfnamefont {B.}~\bibnamefont
  {Sarma}}, \bibinfo {author} {\bibfnamefont {S.}~\bibnamefont {Borah}},
  \bibinfo {author} {\bibfnamefont {A.}~\bibnamefont {Kani}},\ and\ \bibinfo
  {author} {\bibfnamefont {J.}~\bibnamefont {Twamley}},\ }\href
  {https://doi.org/10.1103/PhysRevResearch.4.L042038} {\bibfield  {journal}
  {\bibinfo  {journal} {Phys. Rev. Res.}\ }\textbf {\bibinfo {volume} {4}},\
  \bibinfo {pages} {L042038} (\bibinfo {year} {2022})}\BibitemShut {NoStop}%
\bibitem [{\citenamefont {Goodfellow}\ \emph {et~al.}(2016)\citenamefont
  {Goodfellow}, \citenamefont {Bengio},\ and\ \citenamefont
  {Courville}}]{MIT.Goodfellow.DL}%
  \BibitemOpen
  \bibfield  {author} {\bibinfo {author} {\bibfnamefont {I.}~\bibnamefont
  {Goodfellow}}, \bibinfo {author} {\bibfnamefont {Y.}~\bibnamefont {Bengio}},\
  and\ \bibinfo {author} {\bibfnamefont {A.}~\bibnamefont {Courville}},\
  }\href@noop {} {\emph {\bibinfo {title} {Deep Learning}}}\ (\bibinfo
  {publisher} {MIT Press},\ \bibinfo {year} {2016})\ \bibinfo {note}
  {\url{http://www.deeplearningbook.org}}\BibitemShut {NoStop}%
\bibitem [{\citenamefont {Carleo}\ \emph {et~al.}(2019)\citenamefont {Carleo},
  \citenamefont {Cirac}, \citenamefont {Cranmer}, \citenamefont {Daudet},
  \citenamefont {Schuld}, \citenamefont {Tishby}, \citenamefont
  {Vogt-Maranto},\ and\ \citenamefont {Zdeborov\'a}}]{RevModPhys.91.045002}%
  \BibitemOpen
  \bibfield  {author} {\bibinfo {author} {\bibfnamefont {G.}~\bibnamefont
  {Carleo}}, \bibinfo {author} {\bibfnamefont {I.}~\bibnamefont {Cirac}},
  \bibinfo {author} {\bibfnamefont {K.}~\bibnamefont {Cranmer}}, \bibinfo
  {author} {\bibfnamefont {L.}~\bibnamefont {Daudet}}, \bibinfo {author}
  {\bibfnamefont {M.}~\bibnamefont {Schuld}}, \bibinfo {author} {\bibfnamefont
  {N.}~\bibnamefont {Tishby}}, \bibinfo {author} {\bibfnamefont
  {L.}~\bibnamefont {Vogt-Maranto}},\ and\ \bibinfo {author} {\bibfnamefont
  {L.}~\bibnamefont {Zdeborov\'a}},\ }\href
  {https://doi.org/10.1103/RevModPhys.91.045002} {\bibfield  {journal}
  {\bibinfo  {journal} {Rev. Mod. Phys.}\ }\textbf {\bibinfo {volume} {91}},\
  \bibinfo {pages} {045002} (\bibinfo {year} {2019})}\BibitemShut {NoStop}%
\bibitem [{\citenamefont {Marquardt}(2021)}]{SciPostPhysLectNotes.29}%
  \BibitemOpen
  \bibfield  {author} {\bibinfo {author} {\bibfnamefont {F.}~\bibnamefont
  {Marquardt}},\ }\href {https://doi.org/10.21468/SciPostPhysLectNotes.29}
  {\bibfield  {journal} {\bibinfo  {journal} {SciPost Phys. Lect. Notes}\ ,\
  \bibinfo {pages} {29}} (\bibinfo {year} {2021})}\BibitemShut {NoStop}%
\bibitem [{\citenamefont {Silver}\ \emph {et~al.}(2017)\citenamefont {Silver}
  \emph {et~al.}}]{NatRevPhys.5.141}%
  \BibitemOpen
  \bibfield  {author} {\bibinfo {author} {\bibfnamefont {D.}~\bibnamefont
  {Silver}} \emph {et~al.},\ }\href {https://doi.org/10.1038/nature24270}
  {\bibfield  {journal} {\bibinfo  {journal} {Nature}\ }\textbf {\bibinfo
  {volume} {550}},\ \bibinfo {pages} {354} (\bibinfo {year}
  {2017})}\BibitemShut {NoStop}%
\bibitem [{\citenamefont {Bedolla}\ \emph {et~al.}(2020)\citenamefont
  {Bedolla}, \citenamefont {Padierna},\ and\ \citenamefont
  {Castañeda-Priego}}]{JPhys.33.053001}%
  \BibitemOpen
  \bibfield  {author} {\bibinfo {author} {\bibfnamefont {E.}~\bibnamefont
  {Bedolla}}, \bibinfo {author} {\bibfnamefont {L.~C.}\ \bibnamefont
  {Padierna}},\ and\ \bibinfo {author} {\bibfnamefont {R.}~\bibnamefont
  {Castañeda-Priego}},\ }\href {https://doi.org/10.1088/1361-648X/abb895}
  {\bibfield  {journal} {\bibinfo  {journal} {Journal of Physics: Condensed
  Matter}\ }\textbf {\bibinfo {volume} {33}},\ \bibinfo {pages} {053001}
  (\bibinfo {year} {2020})}\BibitemShut {NoStop}%
\bibitem [{\citenamefont {Carrasquilla}\ and\ \citenamefont
  {Melko}(2017)}]{NatPhys.13.431}%
  \BibitemOpen
  \bibfield  {author} {\bibinfo {author} {\bibfnamefont {J.}~\bibnamefont
  {Carrasquilla}}\ and\ \bibinfo {author} {\bibfnamefont {R.~G.}\ \bibnamefont
  {Melko}},\ }\href@noop {} {\bibfield  {journal} {\bibinfo  {journal} {Nature
  Physics}\ }\textbf {\bibinfo {volume} {13}},\ \bibinfo {pages} {431}
  (\bibinfo {year} {2017})}\BibitemShut {NoStop}%
\bibitem [{\citenamefont {Guest}\ \emph {et~al.}(2018)\citenamefont {Guest},
  \citenamefont {Cranmer},\ and\ \citenamefont
  {Whiteson}}]{AnnuRevNuclPartSci.64.161}%
  \BibitemOpen
  \bibfield  {author} {\bibinfo {author} {\bibfnamefont {D.}~\bibnamefont
  {Guest}}, \bibinfo {author} {\bibfnamefont {K.}~\bibnamefont {Cranmer}},\
  and\ \bibinfo {author} {\bibfnamefont {D.}~\bibnamefont {Whiteson}},\
  }\href@noop {} {\bibfield  {journal} {\bibinfo  {journal} {Annual Review of
  Nuclear and Particle Science}\ }\textbf {\bibinfo {volume} {68}},\ \bibinfo
  {pages} {161} (\bibinfo {year} {2018})}\BibitemShut {NoStop}%
\bibitem [{\citenamefont {George}\ and\ \citenamefont
  {Huerta}(2018)}]{PhysLettB.778.64}%
  \BibitemOpen
  \bibfield  {author} {\bibinfo {author} {\bibfnamefont {D.}~\bibnamefont
  {George}}\ and\ \bibinfo {author} {\bibfnamefont {E.~A.}\ \bibnamefont
  {Huerta}},\ }\href@noop {} {\bibfield  {journal} {\bibinfo  {journal}
  {Physics Letters B}\ }\textbf {\bibinfo {volume} {778}},\ \bibinfo {pages}
  {64} (\bibinfo {year} {2018})}\BibitemShut {NoStop}%
\bibitem [{\citenamefont {Tranter}\ \emph {et~al.}(2018)\citenamefont
  {Tranter}, \citenamefont {Slatyer}, \citenamefont {Hush}, \citenamefont
  {Leung}, \citenamefont {Everett}, \citenamefont {Paul}, \citenamefont
  {Vernaz-Gris}, \citenamefont {Lam}, \citenamefont {Buchler},\ and\
  \citenamefont {Campbell}}]{NatCommun.9.4360}%
  \BibitemOpen
  \bibfield  {author} {\bibinfo {author} {\bibfnamefont {A.~D.}\ \bibnamefont
  {Tranter}}, \bibinfo {author} {\bibfnamefont {H.~J.}\ \bibnamefont
  {Slatyer}}, \bibinfo {author} {\bibfnamefont {M.~R.}\ \bibnamefont {Hush}},
  \bibinfo {author} {\bibfnamefont {A.~C.}\ \bibnamefont {Leung}}, \bibinfo
  {author} {\bibfnamefont {J.~L.}\ \bibnamefont {Everett}}, \bibinfo {author}
  {\bibfnamefont {K.~V.}\ \bibnamefont {Paul}}, \bibinfo {author}
  {\bibfnamefont {P.}~\bibnamefont {Vernaz-Gris}}, \bibinfo {author}
  {\bibfnamefont {P.~K.}\ \bibnamefont {Lam}}, \bibinfo {author} {\bibfnamefont
  {B.~C.}\ \bibnamefont {Buchler}},\ and\ \bibinfo {author} {\bibfnamefont
  {G.~T.}\ \bibnamefont {Campbell}},\ }\href@noop {} {\bibfield  {journal}
  {\bibinfo  {journal} {Nature communications}\ }\textbf {\bibinfo {volume}
  {9}},\ \bibinfo {pages} {4360} (\bibinfo {year} {2018})}\BibitemShut
  {NoStop}%
\bibitem [{\citenamefont {You}\ \emph {et~al.}(2020)\citenamefont {You},
  \citenamefont {Quiroz-Ju{\'a}rez}, \citenamefont {Lambert}, \citenamefont
  {Bhusal}, \citenamefont {Dong}, \citenamefont {Perez-Leija}, \citenamefont
  {Javaid}, \citenamefont {Le{\'o}n-Montiel},\ and\ \citenamefont
  {Maga{\~n}a-Loaiza}}]{ApplPhysRev.7.021404}%
  \BibitemOpen
  \bibfield  {author} {\bibinfo {author} {\bibfnamefont {C.}~\bibnamefont
  {You}}, \bibinfo {author} {\bibfnamefont {M.~A.}\ \bibnamefont
  {Quiroz-Ju{\'a}rez}}, \bibinfo {author} {\bibfnamefont {A.}~\bibnamefont
  {Lambert}}, \bibinfo {author} {\bibfnamefont {N.}~\bibnamefont {Bhusal}},
  \bibinfo {author} {\bibfnamefont {C.}~\bibnamefont {Dong}}, \bibinfo {author}
  {\bibfnamefont {A.}~\bibnamefont {Perez-Leija}}, \bibinfo {author}
  {\bibfnamefont {A.}~\bibnamefont {Javaid}}, \bibinfo {author} {\bibfnamefont
  {R.~d.~J.}\ \bibnamefont {Le{\'o}n-Montiel}},\ and\ \bibinfo {author}
  {\bibfnamefont {O.~S.}\ \bibnamefont {Maga{\~n}a-Loaiza}},\ }\href@noop {}
  {\bibfield  {journal} {\bibinfo  {journal} {Applied Physics Reviews}\
  }\textbf {\bibinfo {volume} {7}},\ \bibinfo {pages} {021404} (\bibinfo {year}
  {2020})}\BibitemShut {NoStop}%
\bibitem [{\citenamefont {Harney}\ \emph {et~al.}(2020)\citenamefont {Harney},
  \citenamefont {Pirandola}, \citenamefont {Ferraro},\ and\ \citenamefont
  {Paternostro}}]{NewJPhys.22.045001}%
  \BibitemOpen
  \bibfield  {author} {\bibinfo {author} {\bibfnamefont {C.}~\bibnamefont
  {Harney}}, \bibinfo {author} {\bibfnamefont {S.}~\bibnamefont {Pirandola}},
  \bibinfo {author} {\bibfnamefont {A.}~\bibnamefont {Ferraro}},\ and\ \bibinfo
  {author} {\bibfnamefont {M.}~\bibnamefont {Paternostro}},\ }\href
  {https://doi.org/10.1088/1367-2630/ab783d} {\bibfield  {journal} {\bibinfo
  {journal} {New Journal of Physics}\ }\textbf {\bibinfo {volume} {22}},\
  \bibinfo {pages} {045001} (\bibinfo {year} {2020})}\BibitemShut {NoStop}%
\bibitem [{\citenamefont {Wise}\ \emph {et~al.}(2021)\citenamefont {Wise},
  \citenamefont {Morton},\ and\ \citenamefont {Dhomkar}}]{PRXQuantum.2.010316}%
  \BibitemOpen
  \bibfield  {author} {\bibinfo {author} {\bibfnamefont {D.~F.}\ \bibnamefont
  {Wise}}, \bibinfo {author} {\bibfnamefont {J.~J.}\ \bibnamefont {Morton}},\
  and\ \bibinfo {author} {\bibfnamefont {S.}~\bibnamefont {Dhomkar}},\ }\href
  {https://doi.org/10.1103/PRXQuantum.2.010316} {\bibfield  {journal} {\bibinfo
   {journal} {PRX Quantum}\ }\textbf {\bibinfo {volume} {2}},\ \bibinfo {pages}
  {010316} (\bibinfo {year} {2021})}\BibitemShut {NoStop}%
\bibitem [{\citenamefont {Vendeiro}\ \emph {et~al.}(2022)\citenamefont
  {Vendeiro}, \citenamefont {Ramette}, \citenamefont {Rudelis}, \citenamefont
  {Chong}, \citenamefont {Sinclair}, \citenamefont {Stewart}, \citenamefont
  {Urvoy},\ and\ \citenamefont {Vuleti\ifmmode~\acute{c}\else
  \'{c}\fi{}}}]{PhysRevResearch.4.043216}%
  \BibitemOpen
  \bibfield  {author} {\bibinfo {author} {\bibfnamefont {Z.}~\bibnamefont
  {Vendeiro}}, \bibinfo {author} {\bibfnamefont {J.}~\bibnamefont {Ramette}},
  \bibinfo {author} {\bibfnamefont {A.}~\bibnamefont {Rudelis}}, \bibinfo
  {author} {\bibfnamefont {M.}~\bibnamefont {Chong}}, \bibinfo {author}
  {\bibfnamefont {J.}~\bibnamefont {Sinclair}}, \bibinfo {author}
  {\bibfnamefont {L.}~\bibnamefont {Stewart}}, \bibinfo {author} {\bibfnamefont
  {A.}~\bibnamefont {Urvoy}},\ and\ \bibinfo {author} {\bibfnamefont
  {V.}~\bibnamefont {Vuleti\ifmmode~\acute{c}\else \'{c}\fi{}}},\ }\href
  {https://doi.org/10.1103/PhysRevResearch.4.043216} {\bibfield  {journal}
  {\bibinfo  {journal} {Phys. Rev. Res.}\ }\textbf {\bibinfo {volume} {4}},\
  \bibinfo {pages} {043216} (\bibinfo {year} {2022})}\BibitemShut {NoStop}%
\bibitem [{\citenamefont {Mohseni}\ \emph {et~al.}(2022)\citenamefont
  {Mohseni}, \citenamefont {F{\"o}sel}, \citenamefont {Guo}, \citenamefont
  {Navarrete-Benlloch},\ and\ \citenamefont {Marquardt}}]{Quantum.6.714}%
  \BibitemOpen
  \bibfield  {author} {\bibinfo {author} {\bibfnamefont {N.}~\bibnamefont
  {Mohseni}}, \bibinfo {author} {\bibfnamefont {T.}~\bibnamefont {F{\"o}sel}},
  \bibinfo {author} {\bibfnamefont {L.}~\bibnamefont {Guo}}, \bibinfo {author}
  {\bibfnamefont {C.}~\bibnamefont {Navarrete-Benlloch}},\ and\ \bibinfo
  {author} {\bibfnamefont {F.}~\bibnamefont {Marquardt}},\ }\href@noop {}
  {\bibfield  {journal} {\bibinfo  {journal} {Quantum}\ }\textbf {\bibinfo
  {volume} {6}},\ \bibinfo {pages} {714} (\bibinfo {year} {2022})}\BibitemShut
  {NoStop}%
\bibitem [{\citenamefont {Ness}\ \emph {et~al.}(2020)\citenamefont {Ness},
  \citenamefont {Vainbaum}, \citenamefont {Shkedrov}, \citenamefont
  {Florshaim},\ and\ \citenamefont {Sagi}}]{PhysRevApplied.14.014011}%
  \BibitemOpen
  \bibfield  {author} {\bibinfo {author} {\bibfnamefont {G.}~\bibnamefont
  {Ness}}, \bibinfo {author} {\bibfnamefont {A.}~\bibnamefont {Vainbaum}},
  \bibinfo {author} {\bibfnamefont {C.}~\bibnamefont {Shkedrov}}, \bibinfo
  {author} {\bibfnamefont {Y.}~\bibnamefont {Florshaim}},\ and\ \bibinfo
  {author} {\bibfnamefont {Y.}~\bibnamefont {Sagi}},\ }\href
  {https://doi.org/10.1103/PhysRevApplied.14.014011} {\bibfield  {journal}
  {\bibinfo  {journal} {Phys. Rev. Appl.}\ }\textbf {\bibinfo {volume} {14}},\
  \bibinfo {pages} {014011} (\bibinfo {year} {2020})}\BibitemShut {NoStop}%
\bibitem [{\citenamefont {Tiunov}\ \emph {et~al.}(2020)\citenamefont {Tiunov},
  \citenamefont {(Vyborova)}, \citenamefont {Ulanov}, \citenamefont {Lvovsky},\
  and\ \citenamefont {Fedorov}}]{Optica.7.448}%
  \BibitemOpen
  \bibfield  {author} {\bibinfo {author} {\bibfnamefont {E.~S.}\ \bibnamefont
  {Tiunov}}, \bibinfo {author} {\bibfnamefont {V.~V.~T.}\ \bibnamefont
  {(Vyborova)}}, \bibinfo {author} {\bibfnamefont {A.~E.}\ \bibnamefont
  {Ulanov}}, \bibinfo {author} {\bibfnamefont {A.~I.}\ \bibnamefont
  {Lvovsky}},\ and\ \bibinfo {author} {\bibfnamefont {A.~K.}\ \bibnamefont
  {Fedorov}},\ }\href {https://doi.org/10.1364/OPTICA.389482} {\bibfield
  {journal} {\bibinfo  {journal} {Optica}\ }\textbf {\bibinfo {volume} {7}},\
  \bibinfo {pages} {448} (\bibinfo {year} {2020})}\BibitemShut {NoStop}%
\bibitem [{\citenamefont {Zeng}\ \emph {et~al.}(2021)\citenamefont {Zeng},
  \citenamefont {Gebremariam}, \citenamefont {Shen}, \citenamefont {Xiong},\
  and\ \citenamefont {Li}}]{ApplPhysLett.118.164003}%
  \BibitemOpen
  \bibfield  {author} {\bibinfo {author} {\bibfnamefont {Y.-X.}\ \bibnamefont
  {Zeng}}, \bibinfo {author} {\bibfnamefont {T.}~\bibnamefont {Gebremariam}},
  \bibinfo {author} {\bibfnamefont {J.}~\bibnamefont {Shen}}, \bibinfo {author}
  {\bibfnamefont {B.}~\bibnamefont {Xiong}},\ and\ \bibinfo {author}
  {\bibfnamefont {C.}~\bibnamefont {Li}},\ }\href@noop {} {\bibfield  {journal}
  {\bibinfo  {journal} {Applied Physics Letters}\ }\textbf {\bibinfo {volume}
  {118}},\ \bibinfo {pages} {164003} (\bibinfo {year} {2021})}\BibitemShut
  {NoStop}%
\bibitem [{\citenamefont {Arrazola}\ \emph {et~al.}(2019)\citenamefont
  {Arrazola}, \citenamefont {Bromley}, \citenamefont {Izaac}, \citenamefont
  {Myers}, \citenamefont {Brádler},\ and\ \citenamefont
  {Killoran}}]{QuantumSciTechnol.4.024004}%
  \BibitemOpen
  \bibfield  {author} {\bibinfo {author} {\bibfnamefont {J.~M.}\ \bibnamefont
  {Arrazola}}, \bibinfo {author} {\bibfnamefont {T.~R.}\ \bibnamefont
  {Bromley}}, \bibinfo {author} {\bibfnamefont {J.}~\bibnamefont {Izaac}},
  \bibinfo {author} {\bibfnamefont {C.~R.}\ \bibnamefont {Myers}}, \bibinfo
  {author} {\bibfnamefont {K.}~\bibnamefont {Brádler}},\ and\ \bibinfo
  {author} {\bibfnamefont {N.}~\bibnamefont {Killoran}},\ }\href
  {https://doi.org/10.1088/2058-9565/aaf59e} {\bibfield  {journal} {\bibinfo
  {journal} {Quantum Science and Technology}\ }\textbf {\bibinfo {volume}
  {4}},\ \bibinfo {pages} {024004} (\bibinfo {year} {2019})}\BibitemShut
  {NoStop}%
\bibitem [{\citenamefont {Ahmed}\ \emph {et~al.}(2021)\citenamefont {Ahmed},
  \citenamefont {S\'anchez Mu\~noz}, \citenamefont {Nori},\ and\ \citenamefont
  {Kockum}}]{PhysRevLett.127.140502}%
  \BibitemOpen
  \bibfield  {author} {\bibinfo {author} {\bibfnamefont {S.}~\bibnamefont
  {Ahmed}}, \bibinfo {author} {\bibfnamefont {C.}~\bibnamefont {S\'anchez
  Mu\~noz}}, \bibinfo {author} {\bibfnamefont {F.}~\bibnamefont {Nori}},\ and\
  \bibinfo {author} {\bibfnamefont {A.~F.}\ \bibnamefont {Kockum}},\ }\href
  {https://doi.org/10.1103/PhysRevLett.127.140502} {\bibfield  {journal}
  {\bibinfo  {journal} {Phys. Rev. Lett.}\ }\textbf {\bibinfo {volume} {127}},\
  \bibinfo {pages} {140502} (\bibinfo {year} {2021})}\BibitemShut {NoStop}%
\bibitem [{\citenamefont {Cantori}\ \emph {et~al.}(2023)\citenamefont
  {Cantori}, \citenamefont {Vitali},\ and\ \citenamefont
  {Pilati}}]{QuantumSciTechnol.8.025022}%
  \BibitemOpen
  \bibfield  {author} {\bibinfo {author} {\bibfnamefont {S.}~\bibnamefont
  {Cantori}}, \bibinfo {author} {\bibfnamefont {D.}~\bibnamefont {Vitali}},\
  and\ \bibinfo {author} {\bibfnamefont {S.}~\bibnamefont {Pilati}},\ }\href
  {https://doi.org/10.1088/2058-9565/acc4e2} {\bibfield  {journal} {\bibinfo
  {journal} {Quantum Science and Technology}\ }\textbf {\bibinfo {volume}
  {8}},\ \bibinfo {pages} {025022} (\bibinfo {year} {2023})}\BibitemShut
  {NoStop}%
\bibitem [{\citenamefont {Bishop}\ and\ \citenamefont
  {Nasrabadi}(2006)}]{Springer.Bishop.PRML}%
  \BibitemOpen
  \bibfield  {author} {\bibinfo {author} {\bibfnamefont {C.~M.}\ \bibnamefont
  {Bishop}}\ and\ \bibinfo {author} {\bibfnamefont {N.~M.}\ \bibnamefont
  {Nasrabadi}},\ }\href@noop {} {\emph {\bibinfo {title} {Pattern recognition
  and machine learning}}},\ Vol.~\bibinfo {volume} {4}\ (\bibinfo  {publisher}
  {Springer},\ \bibinfo {year} {2006})\BibitemShut {NoStop}%
\bibitem [{\citenamefont {Mehta}\ \emph {et~al.}(2019)\citenamefont {Mehta},
  \citenamefont {Bukov}, \citenamefont {Wang}, \citenamefont {Day},
  \citenamefont {Richardson}, \citenamefont {Fisher},\ and\ \citenamefont
  {Schwab}}]{PhysRep.810.1}%
  \BibitemOpen
  \bibfield  {author} {\bibinfo {author} {\bibfnamefont {P.}~\bibnamefont
  {Mehta}}, \bibinfo {author} {\bibfnamefont {M.}~\bibnamefont {Bukov}},
  \bibinfo {author} {\bibfnamefont {C.-H.}\ \bibnamefont {Wang}}, \bibinfo
  {author} {\bibfnamefont {A.~G.}\ \bibnamefont {Day}}, \bibinfo {author}
  {\bibfnamefont {C.}~\bibnamefont {Richardson}}, \bibinfo {author}
  {\bibfnamefont {C.~K.}\ \bibnamefont {Fisher}},\ and\ \bibinfo {author}
  {\bibfnamefont {D.~J.}\ \bibnamefont {Schwab}},\ }\href
  {https://doi.org/https://doi.org/10.1016/j.physrep.2019.03.001} {\bibfield
  {journal} {\bibinfo  {journal} {Physics Reports}\ }\textbf {\bibinfo {volume}
  {810}},\ \bibinfo {pages} {1} (\bibinfo {year} {2019})},\ \bibinfo {note} {a
  high-bias, low-variance introduction to Machine Learning for
  physicists}\BibitemShut {NoStop}%
\bibitem [{\citenamefont {Sutton}\ and\ \citenamefont
  {Barto}(2018)}]{MIT.Sutton.RL}%
  \BibitemOpen
  \bibfield  {author} {\bibinfo {author} {\bibfnamefont {R.~S.}\ \bibnamefont
  {Sutton}}\ and\ \bibinfo {author} {\bibfnamefont {A.~G.}\ \bibnamefont
  {Barto}},\ }\href {https://books.google.co.in/books?id=9jQDxQEACAAJ} {\emph
  {\bibinfo {title} {Reinforcement Learning: An Introduction}}},\ Adaptive
  computation and machine learning\ (\bibinfo  {publisher} {MIT Press},\
  \bibinfo {year} {2018})\BibitemShut {NoStop}%
\bibitem [{\citenamefont {Krenn}\ \emph {et~al.}(2023)\citenamefont {Krenn},
  \citenamefont {Landgraf}, \citenamefont {Foesel},\ and\ \citenamefont
  {Marquardt}}]{PhysRevA.107.010101}%
  \BibitemOpen
  \bibfield  {author} {\bibinfo {author} {\bibfnamefont {M.}~\bibnamefont
  {Krenn}}, \bibinfo {author} {\bibfnamefont {J.}~\bibnamefont {Landgraf}},
  \bibinfo {author} {\bibfnamefont {T.}~\bibnamefont {Foesel}},\ and\ \bibinfo
  {author} {\bibfnamefont {F.}~\bibnamefont {Marquardt}},\ }\href
  {https://doi.org/10.1103/PhysRevA.107.010101} {\bibfield  {journal} {\bibinfo
   {journal} {Phys. Rev. A}\ }\textbf {\bibinfo {volume} {107}},\ \bibinfo
  {pages} {010101} (\bibinfo {year} {2023})}\BibitemShut {NoStop}%
\bibitem [{\citenamefont {Sorokin}\ \emph {et~al.}(2020)\citenamefont
  {Sorokin}, \citenamefont {Ulanov}, \citenamefont {Sazhina},\ and\
  \citenamefont {Lvovsky}}]{NIPS.2020.Interferobot}%
  \BibitemOpen
  \bibfield  {author} {\bibinfo {author} {\bibfnamefont {D.}~\bibnamefont
  {Sorokin}}, \bibinfo {author} {\bibfnamefont {A.}~\bibnamefont {Ulanov}},
  \bibinfo {author} {\bibfnamefont {E.}~\bibnamefont {Sazhina}},\ and\ \bibinfo
  {author} {\bibfnamefont {A.}~\bibnamefont {Lvovsky}},\ }in\ \href
  {https://proceedings.neurips.cc/paper_files/paper/2020/file/99ba5c4097c6b8fef5ed774a1a6714b8-Paper.pdf}
  {\emph {\bibinfo {booktitle} {Advances in Neural Information Processing
  Systems}}},\ Vol.~\bibinfo {volume} {33},\ \bibinfo {editor} {edited by\
  \bibinfo {editor} {\bibfnamefont {H.}~\bibnamefont {Larochelle}}, \bibinfo
  {editor} {\bibfnamefont {M.}~\bibnamefont {Ranzato}}, \bibinfo {editor}
  {\bibfnamefont {R.}~\bibnamefont {Hadsell}}, \bibinfo {editor} {\bibfnamefont
  {M.}~\bibnamefont {Balcan}},\ and\ \bibinfo {editor} {\bibfnamefont
  {H.}~\bibnamefont {Lin}}}\ (\bibinfo  {publisher} {Curran Associates, Inc.},\
  \bibinfo {year} {2020})\ pp.\ \bibinfo {pages} {13238--13248}\BibitemShut
  {NoStop}%
\bibitem [{\citenamefont {Hou}\ \emph {et~al.}(2012)\citenamefont {Hou},
  \citenamefont {Khan}, \citenamefont {Yi}, \citenamefont {Dong},\ and\
  \citenamefont {Petersen}}]{PhysRevA.86.022321}%
  \BibitemOpen
  \bibfield  {author} {\bibinfo {author} {\bibfnamefont {S.~C.}\ \bibnamefont
  {Hou}}, \bibinfo {author} {\bibfnamefont {M.~A.}\ \bibnamefont {Khan}},
  \bibinfo {author} {\bibfnamefont {X.~X.}\ \bibnamefont {Yi}}, \bibinfo
  {author} {\bibfnamefont {D.}~\bibnamefont {Dong}},\ and\ \bibinfo {author}
  {\bibfnamefont {I.~R.}\ \bibnamefont {Petersen}},\ }\href
  {https://doi.org/10.1103/PhysRevA.86.022321} {\bibfield  {journal} {\bibinfo
  {journal} {Phys. Rev. A}\ }\textbf {\bibinfo {volume} {86}},\ \bibinfo
  {pages} {022321} (\bibinfo {year} {2012})}\BibitemShut {NoStop}%
\bibitem [{\citenamefont {O'Driscoll}\ \emph {et~al.}(2019)\citenamefont
  {O'Driscoll}, \citenamefont {Nichols},\ and\ \citenamefont
  {Knott}}]{QuantumMachIntell.1.5}%
  \BibitemOpen
  \bibfield  {author} {\bibinfo {author} {\bibfnamefont {L.}~\bibnamefont
  {O'Driscoll}}, \bibinfo {author} {\bibfnamefont {R.}~\bibnamefont
  {Nichols}},\ and\ \bibinfo {author} {\bibfnamefont {P.~A.}\ \bibnamefont
  {Knott}},\ }\href@noop {} {\bibfield  {journal} {\bibinfo  {journal} {Quantum
  Machine Intelligence}\ }\textbf {\bibinfo {volume} {1}},\ \bibinfo {pages}
  {5} (\bibinfo {year} {2019})}\BibitemShut {NoStop}%
\bibitem [{\citenamefont {Zhang}\ \emph {et~al.}(2020)\citenamefont {Zhang},
  \citenamefont {Zheng}, \citenamefont {Zhang},\ and\ \citenamefont
  {Deng}}]{PhysRevLett.125.170501}%
  \BibitemOpen
  \bibfield  {author} {\bibinfo {author} {\bibfnamefont {Y.-H.}\ \bibnamefont
  {Zhang}}, \bibinfo {author} {\bibfnamefont {P.-L.}\ \bibnamefont {Zheng}},
  \bibinfo {author} {\bibfnamefont {Y.}~\bibnamefont {Zhang}},\ and\ \bibinfo
  {author} {\bibfnamefont {D.-L.}\ \bibnamefont {Deng}},\ }\href
  {https://doi.org/10.1103/PhysRevLett.125.170501} {\bibfield  {journal}
  {\bibinfo  {journal} {Phys. Rev. Lett.}\ }\textbf {\bibinfo {volume} {125}},\
  \bibinfo {pages} {170501} (\bibinfo {year} {2020})}\BibitemShut {NoStop}%
\bibitem [{\citenamefont {Mackeprang}\ \emph {et~al.}(2020)\citenamefont
  {Mackeprang}, \citenamefont {Dasari},\ and\ \citenamefont
  {Wrachtrup}}]{QuantumMachIntell.2.1}%
  \BibitemOpen
  \bibfield  {author} {\bibinfo {author} {\bibfnamefont {J.}~\bibnamefont
  {Mackeprang}}, \bibinfo {author} {\bibfnamefont {D.~B.~R.}\ \bibnamefont
  {Dasari}},\ and\ \bibinfo {author} {\bibfnamefont {J.}~\bibnamefont
  {Wrachtrup}},\ }\href@noop {} {\bibfield  {journal} {\bibinfo  {journal}
  {Quantum Machine Intelligence}\ }\textbf {\bibinfo {volume} {2}},\ \bibinfo
  {pages} {1} (\bibinfo {year} {2020})}\BibitemShut {NoStop}%
\bibitem [{\citenamefont {Metz}\ and\ \citenamefont
  {Bukov}(2023)}]{NatMachIntell.5.780}%
  \BibitemOpen
  \bibfield  {author} {\bibinfo {author} {\bibfnamefont {F.}~\bibnamefont
  {Metz}}\ and\ \bibinfo {author} {\bibfnamefont {M.}~\bibnamefont {Bukov}},\
  }\href@noop {} {\bibfield  {journal} {\bibinfo  {journal} {Nature Machine
  Intelligence}\ }\textbf {\bibinfo {volume} {5}},\ \bibinfo {pages} {780}
  (\bibinfo {year} {2023})}\BibitemShut {NoStop}%
\bibitem [{\citenamefont {F\"osel}\ \emph {et~al.}(2018)\citenamefont
  {F\"osel}, \citenamefont {Tighineanu}, \citenamefont {Weiss},\ and\
  \citenamefont {Marquardt}}]{PhysRevX.8.031084}%
  \BibitemOpen
  \bibfield  {author} {\bibinfo {author} {\bibfnamefont {T.}~\bibnamefont
  {F\"osel}}, \bibinfo {author} {\bibfnamefont {P.}~\bibnamefont {Tighineanu}},
  \bibinfo {author} {\bibfnamefont {T.}~\bibnamefont {Weiss}},\ and\ \bibinfo
  {author} {\bibfnamefont {F.}~\bibnamefont {Marquardt}},\ }\href
  {https://doi.org/10.1103/PhysRevX.8.031084} {\bibfield  {journal} {\bibinfo
  {journal} {Phys. Rev. X}\ }\textbf {\bibinfo {volume} {8}},\ \bibinfo {pages}
  {031084} (\bibinfo {year} {2018})}\BibitemShut {NoStop}%
\bibitem [{\citenamefont {Sommer}\ \emph {et~al.}(2020)\citenamefont {Sommer},
  \citenamefont {Asjad},\ and\ \citenamefont {Genes}}]{SciRep.10.2623}%
  \BibitemOpen
  \bibfield  {author} {\bibinfo {author} {\bibfnamefont {C.}~\bibnamefont
  {Sommer}}, \bibinfo {author} {\bibfnamefont {M.}~\bibnamefont {Asjad}},\ and\
  \bibinfo {author} {\bibfnamefont {C.}~\bibnamefont {Genes}},\ }\href@noop {}
  {\bibfield  {journal} {\bibinfo  {journal} {Scientific Reports}\ }\textbf
  {\bibinfo {volume} {10}},\ \bibinfo {pages} {2623} (\bibinfo {year}
  {2020})}\BibitemShut {NoStop}%
\bibitem [{\citenamefont {Haarnoja}\ \emph {et~al.}(2018)\citenamefont
  {Haarnoja}, \citenamefont {Zhou}, \citenamefont {Abbeel},\ and\ \citenamefont
  {Levine}}]{ICML.2018.SAC}%
  \BibitemOpen
  \bibfield  {author} {\bibinfo {author} {\bibfnamefont {T.}~\bibnamefont
  {Haarnoja}}, \bibinfo {author} {\bibfnamefont {A.}~\bibnamefont {Zhou}},
  \bibinfo {author} {\bibfnamefont {P.}~\bibnamefont {Abbeel}},\ and\ \bibinfo
  {author} {\bibfnamefont {S.}~\bibnamefont {Levine}},\ }in\ \href@noop {}
  {\emph {\bibinfo {booktitle} {Proceedings of the International Conference on
  Machine Learning}}}\ (\bibinfo {year} {2018})\BibitemShut {NoStop}%
\bibitem [{\citenamefont {Kuznetsov}\ \emph {et~al.}(2020)\citenamefont
  {Kuznetsov}, \citenamefont {Shvechikov}, \citenamefont {Grishin},\ and\
  \citenamefont {Vetrov}}]{ICML.2020.TQC}%
  \BibitemOpen
  \bibfield  {author} {\bibinfo {author} {\bibfnamefont {A.}~\bibnamefont
  {Kuznetsov}}, \bibinfo {author} {\bibfnamefont {P.}~\bibnamefont
  {Shvechikov}}, \bibinfo {author} {\bibfnamefont {A.}~\bibnamefont
  {Grishin}},\ and\ \bibinfo {author} {\bibfnamefont {D.}~\bibnamefont
  {Vetrov}},\ }in\ \href@noop {} {\emph {\bibinfo {booktitle} {Proceedings of
  the International Conference on Machine Learning}}}\ (\bibinfo {year}
  {2020})\BibitemShut {NoStop}%
\bibitem [{\citenamefont {Hiraoka}\ \emph {et~al.}(2022)\citenamefont
  {Hiraoka}, \citenamefont {Imagawa}, \citenamefont {Hashimoto}, \citenamefont
  {Onishi},\ and\ \citenamefont {Tsuruoka}}]{ICLR.2022.DroQ}%
  \BibitemOpen
  \bibfield  {author} {\bibinfo {author} {\bibfnamefont {T.}~\bibnamefont
  {Hiraoka}}, \bibinfo {author} {\bibfnamefont {T.}~\bibnamefont {Imagawa}},
  \bibinfo {author} {\bibfnamefont {T.}~\bibnamefont {Hashimoto}}, \bibinfo
  {author} {\bibfnamefont {T.}~\bibnamefont {Onishi}},\ and\ \bibinfo {author}
  {\bibfnamefont {Y.}~\bibnamefont {Tsuruoka}},\ }in\ \href@noop {} {\emph
  {\bibinfo {booktitle} {Proceedings of the International Conference on
  Learning Representations}}}\ (\bibinfo {year} {2022})\BibitemShut {NoStop}%
\bibitem [{\citenamefont {Bhatt}\ \emph {et~al.}(2024)\citenamefont {Bhatt},
  \citenamefont {Palenicek}, \citenamefont {Belousov}, \citenamefont {Argus},
  \citenamefont {Amiranashvili}, \citenamefont {Brox},\ and\ \citenamefont
  {Peters}}]{ICLR.2024.CrossQ}%
  \BibitemOpen
  \bibfield  {author} {\bibinfo {author} {\bibfnamefont {A.}~\bibnamefont
  {Bhatt}}, \bibinfo {author} {\bibfnamefont {D.}~\bibnamefont {Palenicek}},
  \bibinfo {author} {\bibfnamefont {B.}~\bibnamefont {Belousov}}, \bibinfo
  {author} {\bibfnamefont {M.}~\bibnamefont {Argus}}, \bibinfo {author}
  {\bibfnamefont {A.}~\bibnamefont {Amiranashvili}}, \bibinfo {author}
  {\bibfnamefont {T.}~\bibnamefont {Brox}},\ and\ \bibinfo {author}
  {\bibfnamefont {J.}~\bibnamefont {Peters}},\ }in\ \href@noop {} {\emph
  {\bibinfo {booktitle} {Proceedings of the International Conference on
  Learning Representations}}}\ (\bibinfo {year} {2024})\BibitemShut {NoStop}%
\bibitem [{\citenamefont {Bowen}\ and\ \citenamefont
  {Milburn}(2015)}]{TaylorFrancis.QuantumOptomechanics.Bowen}%
  \BibitemOpen
  \bibfield  {author} {\bibinfo {author} {\bibfnamefont {W.~P.}\ \bibnamefont
  {Bowen}}\ and\ \bibinfo {author} {\bibfnamefont {G.~J.}\ \bibnamefont
  {Milburn}},\ }\href {https://books.google.co.in/books?id=xqlcrgEACAAJ} {\emph
  {\bibinfo {title} {Quantum Optomechanics}}}\ (\bibinfo  {publisher} {Taylor
  \& Francis},\ \bibinfo {year} {2015})\BibitemShut {NoStop}%
\bibitem [{\citenamefont {Agarwal}(2013)}]{CUP.QuantumOptics.Agarwal}%
  \BibitemOpen
  \bibfield  {author} {\bibinfo {author} {\bibfnamefont {G.~S.}\ \bibnamefont
  {Agarwal}},\ }\href {https://books.google.co.in/books?id=7KKw_XIYaioC} {\emph
  {\bibinfo {title} {Quantum Optics}}}\ (\bibinfo  {publisher} {Cambridge
  University Press},\ \bibinfo {year} {2013})\BibitemShut {NoStop}%
\bibitem [{sup()}]{suppl}%
  \BibitemOpen
  \href@noop {} {}\bibinfo {note} {See Supplementary document for a detailed
  description on the implmentations and comparisons.}\BibitemShut {Stop}%
\bibitem [{\citenamefont {Bothner}\ \emph {et~al.}(2020)\citenamefont
  {Bothner}, \citenamefont {Yanai}, \citenamefont {Iniguez-Rabago},
  \citenamefont {Yuan}, \citenamefont {Blanter},\ and\ \citenamefont
  {Steele}}]{NatCommun.11.1589}%
  \BibitemOpen
  \bibfield  {author} {\bibinfo {author} {\bibfnamefont {D.}~\bibnamefont
  {Bothner}}, \bibinfo {author} {\bibfnamefont {S.}~\bibnamefont {Yanai}},
  \bibinfo {author} {\bibfnamefont {A.}~\bibnamefont {Iniguez-Rabago}},
  \bibinfo {author} {\bibfnamefont {M.}~\bibnamefont {Yuan}}, \bibinfo {author}
  {\bibfnamefont {Y.~M.}\ \bibnamefont {Blanter}},\ and\ \bibinfo {author}
  {\bibfnamefont {G.~A.}\ \bibnamefont {Steele}},\ }\href
  {https://doi.org/10.1038/s41467-020-15389-4} {\bibfield  {journal} {\bibinfo
  {journal} {Nat. Commun.}\ }\textbf {\bibinfo {volume} {11}},\ \bibinfo
  {pages} {1589} (\bibinfo {year} {2020})}\BibitemShut {NoStop}%
\bibitem [{\citenamefont {Asjad}\ \emph {et~al.}(2014)\citenamefont {Asjad},
  \citenamefont {Agarwal}, \citenamefont {Kim}, \citenamefont {Tombesi},
  \citenamefont {Giuseppe},\ and\ \citenamefont {Vitali}}]{PhysRevA.89.023849}%
  \BibitemOpen
  \bibfield  {author} {\bibinfo {author} {\bibfnamefont {M.}~\bibnamefont
  {Asjad}}, \bibinfo {author} {\bibfnamefont {G.~S.}\ \bibnamefont {Agarwal}},
  \bibinfo {author} {\bibfnamefont {M.~S.}\ \bibnamefont {Kim}}, \bibinfo
  {author} {\bibfnamefont {P.}~\bibnamefont {Tombesi}}, \bibinfo {author}
  {\bibfnamefont {G.~D.}\ \bibnamefont {Giuseppe}},\ and\ \bibinfo {author}
  {\bibfnamefont {D.}~\bibnamefont {Vitali}},\ }\href
  {https://doi.org/10.1103/PhysRevA.89.023849} {\bibfield  {journal} {\bibinfo
  {journal} {Phys. Rev. A}\ }\textbf {\bibinfo {volume} {89}},\ \bibinfo
  {pages} {023849} (\bibinfo {year} {2014})}\BibitemShut {NoStop}%
\bibitem [{\citenamefont {Heinrich}\ \emph {et~al.}(2011)\citenamefont
  {Heinrich}, \citenamefont {Ludwig}, \citenamefont {Qian}, \citenamefont
  {Kubala},\ and\ \citenamefont {Marquardt}}]{PhysRevLett.107.043603}%
  \BibitemOpen
  \bibfield  {author} {\bibinfo {author} {\bibfnamefont {G.}~\bibnamefont
  {Heinrich}}, \bibinfo {author} {\bibfnamefont {M.}~\bibnamefont {Ludwig}},
  \bibinfo {author} {\bibfnamefont {J.}~\bibnamefont {Qian}}, \bibinfo {author}
  {\bibfnamefont {B.}~\bibnamefont {Kubala}},\ and\ \bibinfo {author}
  {\bibfnamefont {F.}~\bibnamefont {Marquardt}},\ }\href
  {https://doi.org/10.1103/PhysRevLett.107.043603} {\bibfield  {journal}
  {\bibinfo  {journal} {Phys. Rev. Lett.}\ }\textbf {\bibinfo {volume} {107}},\
  \bibinfo {pages} {043603} (\bibinfo {year} {2011})}\BibitemShut {NoStop}%
\bibitem [{\citenamefont {Ludwig}\ and\ \citenamefont
  {Marquardt}(2013)}]{PhysRevLett.111.073603}%
  \BibitemOpen
  \bibfield  {author} {\bibinfo {author} {\bibfnamefont {M.}~\bibnamefont
  {Ludwig}}\ and\ \bibinfo {author} {\bibfnamefont {F.}~\bibnamefont
  {Marquardt}},\ }\href {https://doi.org/10.1103/PhysRevLett.111.073603}
  {\bibfield  {journal} {\bibinfo  {journal} {Phys. Rev. Lett.}\ }\textbf
  {\bibinfo {volume} {111}},\ \bibinfo {pages} {073603} (\bibinfo {year}
  {2013})}\BibitemShut {NoStop}%
\bibitem [{\citenamefont {Brun}(2002)}]{AmJPhys.70.719}%
  \BibitemOpen
  \bibfield  {author} {\bibinfo {author} {\bibfnamefont {T.~A.}\ \bibnamefont
  {Brun}},\ }\href {https://doi.org/10.1119/1.1475328} {\bibfield  {journal}
  {\bibinfo  {journal} {American Journal of Physics}\ }\textbf {\bibinfo
  {volume} {70}},\ \bibinfo {pages} {719} (\bibinfo {year} {2002})}\BibitemShut
  {NoStop}%
\bibitem [{\citenamefont {Johansson}\ \emph {et~al.}(2012)\citenamefont
  {Johansson}, \citenamefont {Nation},\ and\ \citenamefont
  {Nori}}]{ComputPhysCommun.183.1760}%
  \BibitemOpen
  \bibfield  {author} {\bibinfo {author} {\bibfnamefont {J.~R.}\ \bibnamefont
  {Johansson}}, \bibinfo {author} {\bibfnamefont {P.~D.}\ \bibnamefont
  {Nation}},\ and\ \bibinfo {author} {\bibfnamefont {F.}~\bibnamefont {Nori}},\
  }\href {https://doi.org/10.1016/j.cpc.2012.02.021} {\bibfield  {journal}
  {\bibinfo  {journal} {Comput. Phys. Commun}\ }\textbf {\bibinfo {volume}
  {183}},\ \bibinfo {pages} {1760} (\bibinfo {year} {2012})}\BibitemShut
  {NoStop}%
\bibitem [{\citenamefont {Johansson}\ \emph {et~al.}(2013)\citenamefont
  {Johansson}, \citenamefont {Nation},\ and\ \citenamefont
  {Nori}}]{ComputPhysCommun.184.1234}%
  \BibitemOpen
  \bibfield  {author} {\bibinfo {author} {\bibfnamefont {J.~R.}\ \bibnamefont
  {Johansson}}, \bibinfo {author} {\bibfnamefont {P.~D.}\ \bibnamefont
  {Nation}},\ and\ \bibinfo {author} {\bibfnamefont {F.}~\bibnamefont {Nori}},\
  }\href {https://doi.org/10.1016/j.cpc.2012.11.019} {\bibfield  {journal}
  {\bibinfo  {journal} {Comput. Phys. Commun}\ }\textbf {\bibinfo {volume}
  {184}},\ \bibinfo {pages} {1234} (\bibinfo {year} {2013})}\BibitemShut
  {NoStop}%
\bibitem [{qua()}]{quantrl}%
  \BibitemOpen
  \href@noop {} {}\bibinfo {note} {This work uses our multi-backend numerical
  framework \href{https://github.com/Sampreet/cool_cluster}{\texttt{quantrl}}
  and interfaces libraries like
  \href{https://github.com/qutip/qutip}{\texttt{qutip}},
  \href{https://github.com/qutip/qutip-jax}{\texttt{qutip-jax}},
  \href{https://github.com/DLR-RM/stable-baselines3}{\texttt{stable-baselines3}}
  and \href{https://github.com/araffin/sbx}{\texttt{sbx}}.}\BibitemShut {Stop}%
\end{thebibliography}%


\begin{thebibliography}{10}

\bibitem{MIT.Sutton.RL}
Sutton, R.~S. and Barto, A.~G.
\newblock {\em Reinforcement Learning: An Introduction}.
\newblock Adaptive computation and machine learning. MIT Press,  (2018).

\bibitem{ICML.2018.SAC}
Haarnoja, T., Zhou, A., Abbeel, P., and Levine, S.
\newblock In {\em Proceedings of the International Conference on Machine
  Learning},  (2018).

\bibitem{ICML.2020.TQC}
Kuznetsov, A., Shvechikov, P., Grishin, A., and Vetrov, D.
\newblock In {\em Proceedings of the International Conference on Machine
  Learning},  (2020).

\bibitem{ICLR.2022.DroQ}
Hiraoka, T., Imagawa, T., Hashimoto, T., Onishi, T., and Tsuruoka, Y.
\newblock In {\em Proceedings of the International Conference on Learning
  Representations},  (2022).

\bibitem{ICLR.2024.CrossQ}
Bhatt, A., Palenicek, D., Belousov, B., Argus, M., Amiranashvili, A., Brox, T.,
  and Peters, J.
\newblock In {\em Proceedings of the International Conference on Learning
  Representations},  (2024).

\bibitem{ICLR.2021.REDQ}
Chen, X., Wang, C., Zhou, Z., and Ross, K.
\newblock In {\em Proceedings of the International Conference on Learning
  Representations},  (2021).

\bibitem{JMachLearnRes.22.1}
Raffin, A., Hill, A., Gleave, A., Kanervisto, A., Ernestus, M., and Dormann, N.
\newblock {\em Journal of Machine Learning Research}{ \bf 22}(268), 1--8
  (2021).

\bibitem{PhysRevA.107.053521}
Ghosh, A., Kumar, P., Sommer, C., Jimenez, F.~G., Sudhir, V., and Genes, C.
\newblock {\em Phys. Rev. A}{ \bf 107}, 053521 May  (2023).

\bibitem{PhysRevResearch.4.L042038}
Sarma, B., Borah, S., Kani, A., and Twamley, J.
\newblock {\em Phys. Rev. Res.}{ \bf 4}, L042038 Nov  (2022).

\bibitem{rl-zoo3}
Raffin, A.
\newblock \url{https://github.com/DLR-RM/rl-baselines3-zoo},  (2020).

\end{thebibliography}

\end{document}


\title{Supplementary for Domino-cooling Oscillator Networks with Deep Reinforcement Learning}
    \author{Sampreet Kalita and Amarendra K. Sarma}
    \date{August 15, 2024}

    \maketitle

    \section*{The Reinforcement Learning Framework}
        Unlike supervised and unsupervised learning, the paradigm of reinforcement learning (RL) has only displayed a rapid growth from the mid 2010s.
        A typical RL setup relies on a software or a hardware agent that observes and acts on an (a partially observable) environment to evolve it towards a goal \cite{MIT.Sutton.RL}.
        The reward recieved for the actions at each step are utilized to update the RL-agent's policy and steer the environment's dynamics towards the goal.
        In our work, we utilize the actor-critic based approaches of RL \cite{ICML.2018.SAC, ICML.2020.TQC, ICLR.2022.DroQ, ICLR.2024.CrossQ} to optimize the policy of an actor network which is regulated by a critic network that evaluates the action's value.
        Specifically, we implement the vanilla soft actor-critic (SAC) algorithm \cite{ICML.2018.SAC} and its derivatives \textemdash{} truncated quantile critics (TQC) \cite{ICML.2020.TQC} and the current state-of-the-art CrossQ \cite{ICLR.2024.CrossQ} \textemdash{} to obtain the optimal policies for our models.
        Further, we utilize the configuration of dropout Q functions (DroQ) \cite{ICLR.2022.DroQ} alongside SAC/TQC.

        SAC \cite{ICML.2018.SAC} combines the RL's policy-gradient and value-based approaches with entropy regularization to favour exploration for off-policy deep RL.
        To briefly illustrate this, let us denote the actual state of the environment at time $t$ as $s_{t}$, its observed state as $o_{t}$ and the actions as $a_{t}$.
        These actions are sampled from a probability distribution that determines the policy $\pi (a | o)$.
        The observed states, actions, and rewards ($r_{t + 1} = r (o_{t}, a_{t}, o_{t + 1})$), obtained from the state transitions due to the actions, are then accumulated over the complete trajectory $\tau = ( o_{0}, a_{0}, o_{1}, a_{1}, \dots, o_{T - 1}, a_{T - 1}, o_{T} )$, such that the return at time $t$ is given by $R_{t + 1} = \Sigma_{t^{\prime} = 0}^{T - t - 1} \gamma^{t^{\prime}} r_{t + t^{\prime} + 1}$ with discount factor $\gamma$.
        The agent learns to maximize both the return and the entropy $\mathcal{H} (\pi) = \mathbb{E} [ - \ln \pi]$ to obtain the optimal policy
        \begin{eqnarray}
            \pi^{*} = \arg\max \sum_{t = 0}^{T - 1} \gamma^{t} \mathbb{E}_{( s_{t}, a_{t} ) \sim \rho_{\pi}} \left[ r_{t + 1} + \alpha \mathcal{H} (\cdot | o_{t}) \right],
        \end{eqnarray}
        where $\rho_{\pi}$ denotes the state-action marginal of the trajectory and $\alpha$ is the temperature.
        The policy parameters are updated by minimizing the expected Kullback-Leibler (KL) divergence ($D_{KL}$) \cite{MIT.Sutton.RL} and performing gradient descent over
        \begin{eqnarray}
            J_{\pi} = \mathbb{E} \left[ D_{KL} \left( \pi || \frac{\exp [ Q (o_{t}, \cdot)]}{Z} \right) \right],
        \end{eqnarray}
        where $Z (o_{t})$ is the partition function.
        Here, $Q (o_{t}, a_{t})$ denotes the action-value and is optimized by minimizing the the soft Bellman residual
        \begin{eqnarray}
            J_{Q} = \mathbb{E}_{( s_{t}, a_{t} ) \sim \mathcal{D}} \left[ \left( Q_{\pi} - Q_{\pi}^{\prime} \right)^{2} \right],
        \end{eqnarray}
        with $Q_{\pi}^{\prime}$ updated by using the Bellman equation $Q_{\pi}^{\prime} = r_{t + 1} + \gamma \mathbb{E}_{o_{t + 1} \sim p} \left[ V_{\pi} (o_{t + 1}) \right]$, where $p$ denotes the transition probability from state $s_{t}$ and $\mathcal{D}$ is the off-policy distribution.
        The value function $V_{\pi} (o_{t})$ is related to the Q-function as
        \begin{eqnarray}
            V_{\pi} = \mathbb{E}_{o_{t + 1} \sim p} \left[ Q_{\pi} + \alpha \mathcal{H} (\pi) \right],
        \end{eqnarray}
        and is updated by minimizing the squared residual error
        \begin{eqnarray}
            J_{V} = \mathbb{E}_{s_{t} \sim \mathcal{D}} \left[ \frac{1}{2} \left( V_{\pi} - \mathbb{E}_{a_{t} \sim \pi} \left[ Q_{\pi} - \ln \pi \right] \right)^{2} \right].
        \end{eqnarray}

\begin{figure}[!h]
            \centering
            \includegraphics[width=0.84\textwidth]{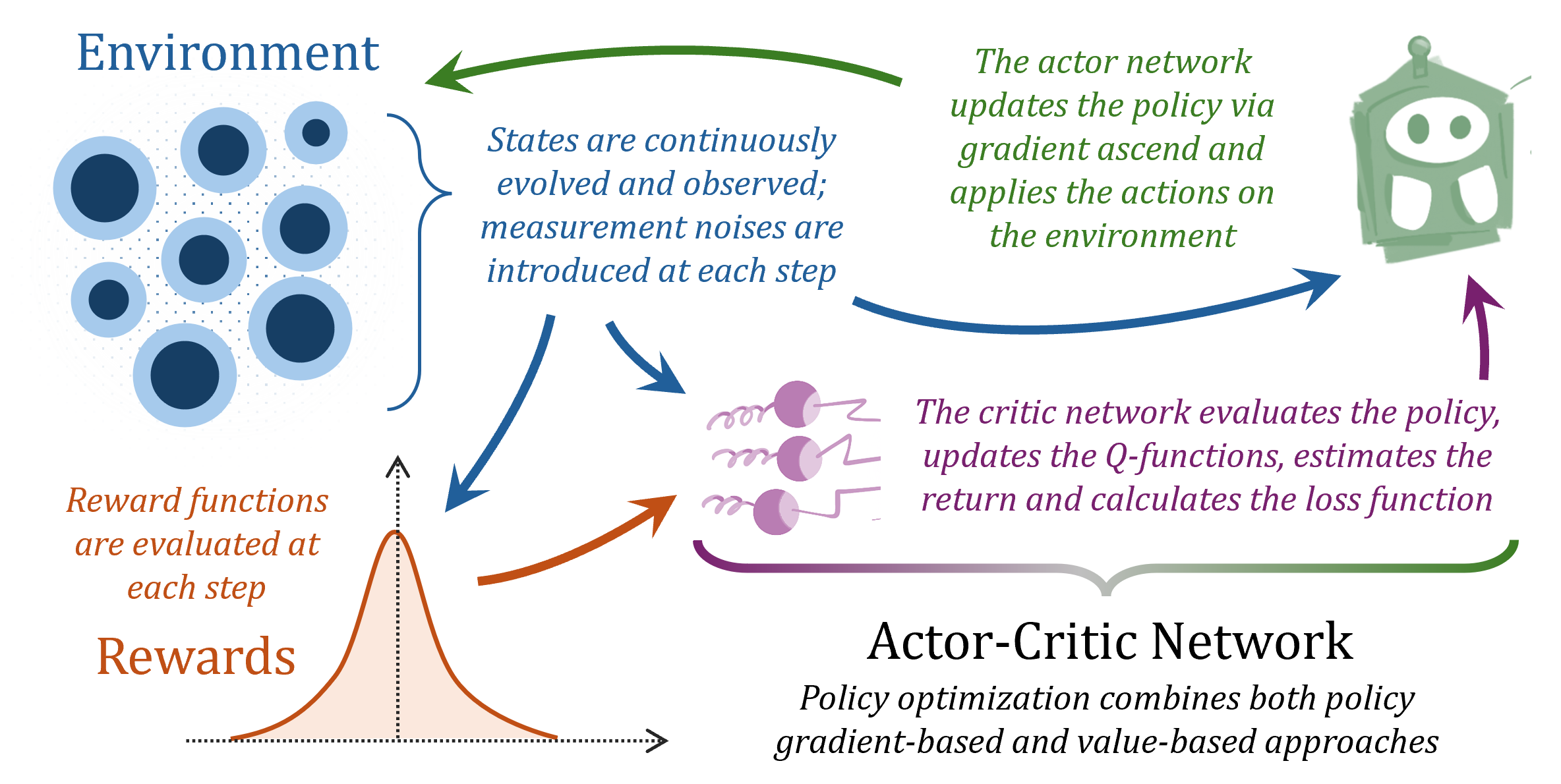}
            \caption{
                \textit{Framework:} RL-based approach to cool down harmonic oscillators.
            }
            \label{fig:s0}
        \end{figure}

        Whereas vanilla SAC uses the minimum of a collection of Q-functions for the critic networks for entropy regularization, TQC implements a distributional representation of the Q-functions to randomize the return and introduces granularity to control the overestimation \cite{ICML.2020.TQC}.
        Moreover, it truncates the topmonst atoms of the distribution in the ensemble and results in improved performance.
        It was also shown in later works that increasing the number of gradient steps improved the sample efficiency of SAC \cite{ICLR.2021.REDQ}, and implementing dropout layers in the Q-function network over it further improved the performance \cite{ICLR.2022.DroQ}.
        However, these implementations increased the computational cost several folds.
        This issue was recently resolved by the addition of batch normalization and the removal of target networks in CrossQ \cite{ICLR.2024.CrossQ}.
        For our work, we use our library \texttt{quantrl} to interface different harmonic oscillator environments and train the RL-agent using the JAX implementation of SAC/TQC/CrossQ available in \texttt{stable-baseline3} \cite{JMachLearnRes.22.1}.
        A pictorial depiction of the implementation and the training is given in Fig. \ref{fig:s0}.
        
        \newpage
    
    \section*{Cooling Independent Harmonic Oscillators}
        We first evaluate our RL-based approach for the feedback-cooling of \textit{independent} harmonic oscillators.
        The effective Hamiltonian of the $j$-th oscillator is given by $H_{j} = \hbar \Omega_{j} ( q_{j}^{2} + p_{j}^{2} ) / 2 - \hbar \eta_{j} q_{j}^{2}$, where $\Omega_{j}$, $q_{j}$ and $p_{j}$ are the frequency, position and momentum of the oscillator, and $\eta_{j}$ is the strength of the external feedback.
        It can be analytically shown \cite{PhysRevA.107.053521} that by applying modulated forces of strength $\eta_{j} = \eta \cos{( 2 \Omega_{j} t + \phi_{j}^{*} )}$ with \textit{phase kicks} $\phi_{j}^{*} = \pi / 2 + \tanh{[ 1 / \{ p_{j} / q_{j} + \eta / ( 2 \Omega_{j} ) \} ]}$, the oscillators can be successively cooled down to an occupancy $\sim \gamma_{j} n_{th} / \eta$, where $\gamma_{j}$ and $n_{th}$ denote the decay rate of the oscillator and the mean thermal occupancy respectively. 
        Here, we consider a more general form of the strength, $\eta_{j} = \eta \cos{( 2 \Omega t + \phi_{j} )}$, to cool down oscillators whose frequencies lie \textit{around} $\Omega$.
        To obtain the dynamics of oscillators in the presence of this modulated feedback, we perform Wiener increments as detailed in the main text.
        Finally, we interface the RL-agent by feeding it with the position and momentum values periodically and by evolving the oscillators using the phase kicks predicted by it.
        The agent is then rewarded on the basis of the amount of cooling achieved at each step.

        \begin{figure}[!h]
            \centering
            \includegraphics[width=0.96\textwidth]{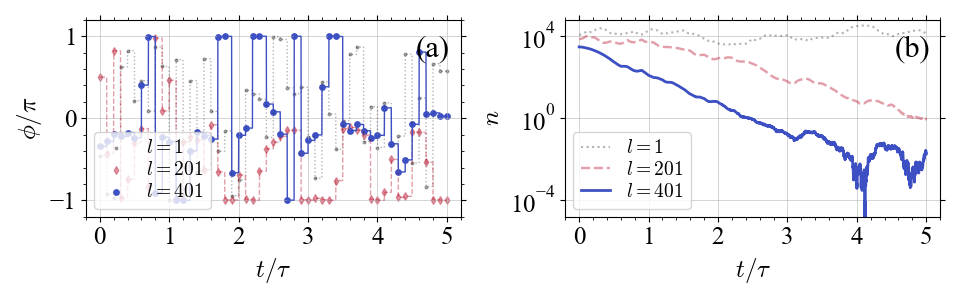}
            \caption{
                \textit{Cooling independent oscillators:}
                (a) Phase kicks ($\phi$) learned by the RL-agent in the $l$-th training episode and (b) dimensionless thermal energies ($n$) obtained with these phase kicks.
                The parameters used are $d \Omega = 0.1 \Omega$, $\eta = 0.1 \Omega$, $\gamma_{j} = 10^{-6} \Omega$ and $n_{th} = 10^{4}$, with $\tau = \Omega^{-1}$.
                The kicks are imparted every $0.1 \tau$.
            }
            \label{fig:s1}
        \end{figure}

        In Fig. \ref{fig:s1}(a), we plot the sequence of phase kicks predicted by our RL-agent during different stages of training.
        The corresponding dynamics of normalized thermal energies $n_{j} = ( q_{j}^{2} + p_{j}^{2} ) / 2$ is shown in Fig. \ref{fig:s1}(b).
        We observe that after a few hundred episodes, the agent successfully predicts the optimal phase kicks to successively cool every oscillator close to $\gamma_{j} n_{th} / \eta$, where the effect of feedback-cooling is comparable to that of the Brownian noise.
        Here, it may be important to note that each episode starts from a unique initial state: the frequency of the oscillator is sampled from a normal distribution centered at $\Omega$ with a standard deviation of $d \Omega$, and its position and momentum are obtained from a thermal distribution of mean occupancy $n_{th}$, such that they obey $\langle q_{j}^{2} \rangle = \langle p_{j}^{2} \rangle = n_{th} + 1 / 2$.
        This is done to make the training robust towards different oscillator frequencies, resulting in a generalized model for the prediction of optimal phase kicks.
        
\begin{figure}[!ht]
            \centering
            \includegraphics[width=0.96\textwidth]{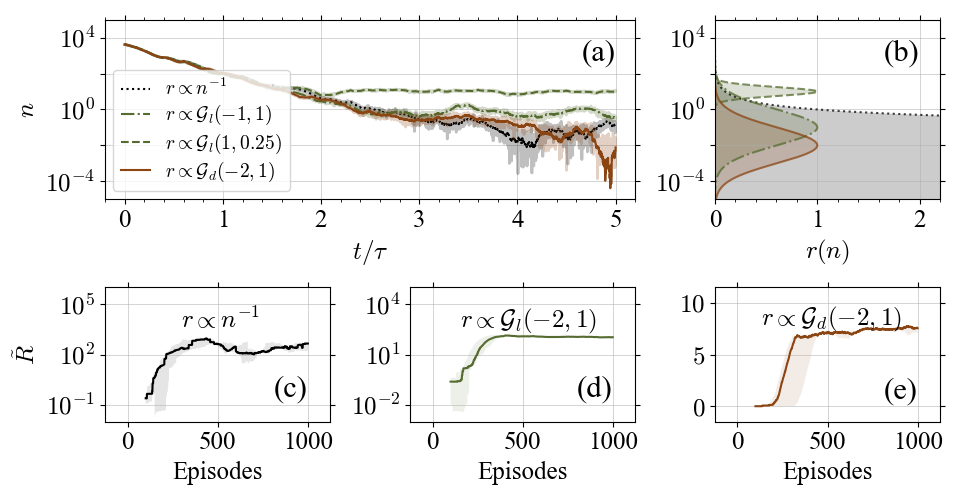}
            \caption{
                \textit{Selecting the reward function:}
                (a) Time evolution of trajectories with the same initial conditions but different types of reward functions obtained after training each of them for $10^{3}$ episodes with the same seed.
                (b) Illustration of the rewards corresponding to each of the functions for different values of $n$.
                (c-e) Comparison of interquartile ranges ($10^{3}$ trajectories over $5$ different seeds) of the mean episode reward per $100$ episodes ($\tilde{R}$) for all three approaches.
                Here, the measurement imprecisions are zero-mean Gaussian noises with a standard deviation of $\sigma_{m} = 0.1$.
                Other parameters used are the same as Fig. \ref{fig:s1}.
            }
            \label{fig:s2}
        \end{figure}

        We now take into account the imprecision in the measurement of the position and momentum of the oscillators.
        Since the optimal values of the phase kicks are highly sensitive to these measured values, the time evolution significantly deviates from an ideal trajectory as the measurements tend to grown noisier.
        However, using our RL-based approach, we show that one can still achieve a degree of cooling similar to the noiseless scenario, as depicted in Fig. \ref{fig:s2}(a).
        Here, the Gaussian noise introduced into the measurements are visible as fluctuations in the observed thermal occupancies.
        It may also be noted here that in order to obtain higher rewards for lower values of occupancies, we initially considered a reward function of the form $r_{j} \propto 1 / n_{j}$ \cite{PhysRevResearch.4.L042038}.
        However, this function is unbounded and results in steep gradients as the occupancies go below unity.
        To circumvent this, we introduce a Gaussian form of the reward function as $r_{j} \propto \mathcal{G}_{l} (\mu, \sigma) = \exp[ - ( \log_{10} [ n_{j} ] - \mu )^{2} / 2 \sigma^{2}]$, such that the target occupancy and the standard deviations are $10^{\mu}$ and $\sigma$ respectively.
        This results in three important outcomes: (i) the reward is smoothly bounded in the closed interval $(0, 1]$, (ii) the degree of cooling can be externally controlled since the maximum value of this reward function peaks at the target occupancy, and (iii) the degree of relaxation from the target occupancy can be controlled by tuning the standard deviation value.
        These points are illustrated in Fig. \ref{fig:s2}(b).
        This scheme may be further improved by using a difference function of the form $r_{j} \propto \mathcal{G}_{d} (\mu, \sigma) = \mathcal{G}_{l}^{(t + 1)} - \mathcal{G}_{l}^{(t)}$.
        We plot the learning curves for each of these reward functions in Fig. \ref{fig:s2}(c-e).
        It can be seen that the difference function leads to a significant enhancement in training.

        Once the RL-agent has been trained to obtain the optimal strategy to cool any harmonic oscillator, it can be vectorized to simultaneously cool a collection of independent harmonic oscillators.

    \section*{Cooling Coupled Harmonic Oscillators}
        For linearly coupled harmonic oscillators whose system Hamiltonian is given by Eq. (1) of the main text, the analytical expression for phase-adaptive feedback may not be able to simultaneously cool all the participating oscillators, specially for higher values of $N$.
        We therefore train the RL-agent to learn the sequence of individual phase-kicks imparted \textit{only} to the leaf nodes, using \textit{all} the measured values of positions and momentums.
        This introduces a second layer of feedback-coupling between the oscillators \cite{PhysRevA.107.053521} which is utilized by the RL-agent to steer the dynamics towards cooling.
        In what follows, we first describe the formalism used for the $N$-oscillator networks and then discuss the results for feedback-cooling the internal nodes using the leaf nodes in multiple configurations.

        A minimally connected graph with $N$ nodes requires $N - 1$ edges and therefore, denotes a tree, where each node $j \in [1, N]$ is connected to $e_{k}$ ($k \in [1, N - 1]$) other nodes.
        To create such a tree, one can use the Pr\"{u}fer sequence to generate a set of edges $E_{k}$ such that the $k$-th edge connects $j_{k}$ and $j^{\prime}_{k}$.
        However, for our specific case-studies, we select the configurations manually.
        Then, at each node, we place an oscillator, which is connected to the other oscillators of the network either directly or indirectly.
        We then implement the time evolution by following the Wiener increment process described in the main text.
        Finally, the RL-agent is trained to predict the optimal values of $\phi_{j}$ in the external feedback forces on the leaf nodes, by maximizing the cumulative reward at the end of each episode, based the the target occupancies $\mu_{j}$ and relaxation $\sigma_{j}$ provided for each node.

        \begin{figure}[!h]
            \centering
            \includegraphics[width=0.96\textwidth]{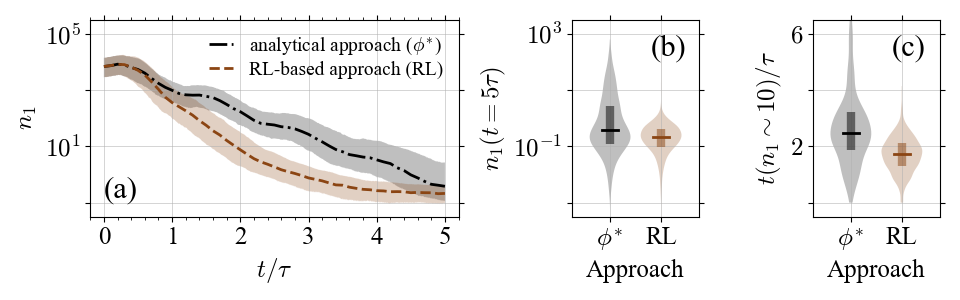}
            \caption{
                \textit{Cooling coupled oscillators:}
                (a) Interquartile ranges of dimensionless energies $n_{1}$ for the central ($j = 1$) oscillator in a configuration with $N = 3$ with two leaves ($j = 2, 3$), obtained with analytical (dash-dotted black line) and RL-based (dashed brown line) approaches for $10^{3}$ trajectories sampled from a thermal occupancy of $n_{th} = 10^{4}$.
                The corresponding interquartile ranges for energies at $t = 5 \tau$ and time required reach $n \sim 10$ are depicted in (b) and (c) respectively.
                Parameters used are the same as Fig. 3 of the main text.
            }
            \label{fig:s3}
        \end{figure}

\begin{figure}[!h]
            \centering
            \includegraphics[width=0.96\textwidth]{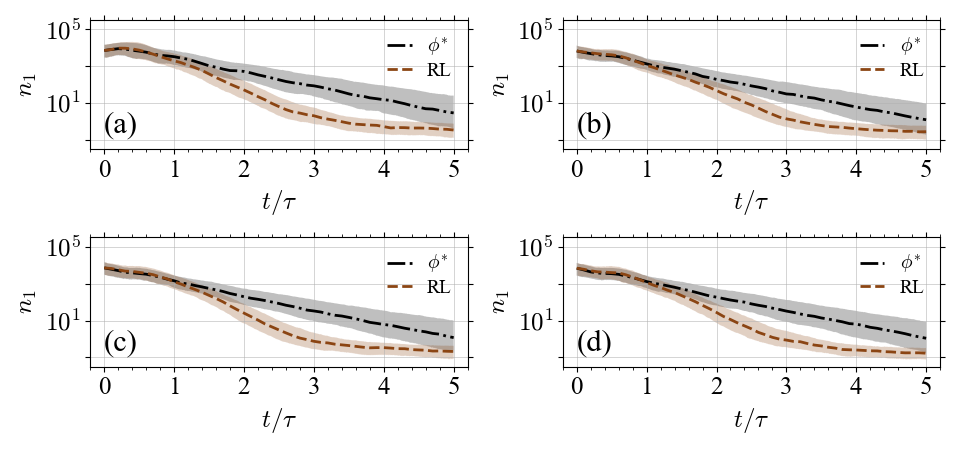}
            \caption{
                \textit{Cooling 4 coupled oscillators with 3 leaves and 1 internal oscillator:}
                Here, $j = 1$ denotes the internal oscillator.
                Other parameters are as Fig. \ref{fig:s3}(a).
            }
            \label{fig:s4}
        \end{figure}

\begin{figure}[!h]
            \centering
            \includegraphics[width=0.96\textwidth]{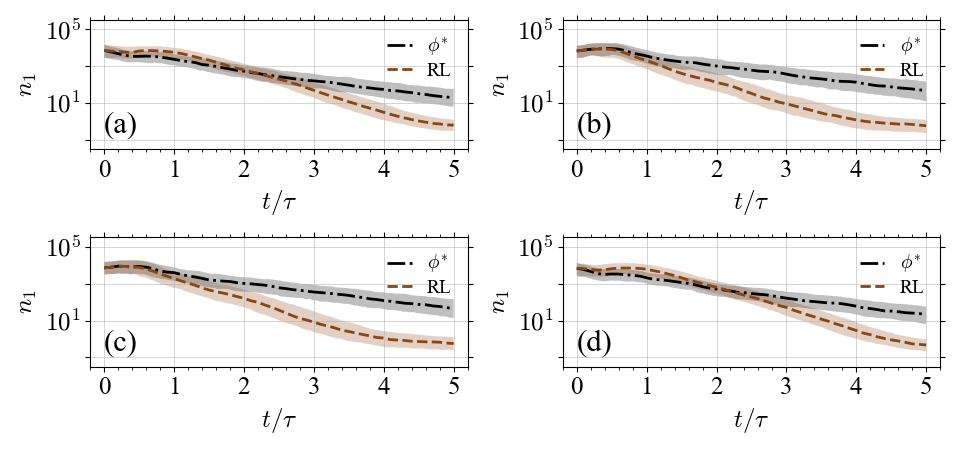}
            \caption{
                \textit{Cooling 4 coupled oscillators with 2 leaves and 2 internal oscillators:}
                Here, $j = 2, 3$ denote the internal oscillators.
                Other parameters are as Fig. \ref{fig:s3}(a).
            }
            \label{fig:s5}
        \end{figure}

        Here, we briefly demonstrate our approach with 3 and 4 coupled oscillators.
        For $N = 3$, we impart external feedback only to the two leaf-node oscillators.
        It can be seen from Fig. \ref{fig:s3} that, compared to the scenario of cascaded cooling with the optimal phase obtained from the analytical expression, the rate of cooling in the RL-based approach is significantly higher.
        We observe similar outcomes for $N = 4$ with 3-leaf and 2-leaf configurations (refer to Fig. 1 of the main text), as depicted in Fig. \ref{fig:s4} and Fig. \ref{fig:s5} respectively.
        Proceeding in a similar way, we increase the total number of oscillators upto $N = 9$ and report our observations for some special configurations with fixed frequencies in the main text.

    \section*{A Note on the Fully Quantum Treatment}
        For the semi-classical Wiener process, we used the measurements of $q_{j}$ and $p_{j}$ to predict the feedback phase kicks from the RL-agent in order to optimize the dimensionless energy $n_{j}$.
        However, the measurement of observables in a fully quantum treatment collapses the state of the system to one of the eigenstates of the observable.
        Owing to the time complexity for the simulation of Monte-Carlo quantum trajectories with periodic measurements and known limitations of the measurements for higher dimensional Hilbert spaces, we resort to calculating the expectation values of $q_{j}^{2}$, $p_{j}^{2}$ and $q_{j} p_{j}$ and estimate $\langle n_{j} \rangle = ( \langle q_{2} \rangle + \langle p_{2} \rangle - 1 ) / 2$ for the rewards.
        We further restrict our analysis to independent oscillators and show that the RL-agent successfully cools the oscillators far below the ensembled average limit of $\langle n_{j} \rangle \sim \gamma_{j} n_{th} / \eta$.

    \section*{Ablation Study}
        Table \ref{tab:1} presents the hyperparameters that worked best for our simulations.
        We observe that the number of hidden layers significantly affected the ability to cool the oscillators.
        Two other important hyperparameters in our work are the scale of the reward and the entropy coefficient.
        We have also varied the learning rates of the policy gradients and Q functions.
        Other parameters are the same as those deemed optimal for the corresponding algorithms \cite{rl-zoo3}.

        \begin{table}[!h]
            \centering
            \caption{
                Hyperparamters used for the simulations.
            }
            \label{tab:1}
            \begin{tabular}{|l||c|c|c|}
                \hline
                \textbf{Hyperparamter} & \multicolumn{2}{c|}{\textbf{SAC / TQC}} & \textbf{CrossQ}\\
                \hline
                Policy Learning Rate & \multicolumn{3}{c|}{0.0003 ($N \leq 4$), 0.001 ($N > 4$)} \\
                \hline
                Q-function Learning Rate & \multicolumn{2}{c|}{0.001} & 0.0003, 0.001 \\
                \hline
                Replay Buffer Size & \multicolumn{3}{c|}{1,000,000} \\
                \hline
                Batch Size & \multicolumn{3}{c|}{512} \\
                \hline
                Discount Factor & \multicolumn{3}{c|}{0.95} \\
                \hline
                Entropy Coefficient & \multicolumn{3}{c|}{0.01} \\
                \hline
                Number of Layers & \multicolumn{3}{c|}{3 ($N < 4$), 4 ($N \geq 4$)} \\
                \hline
                Actor Layer Width & \multicolumn{3}{c|}{256} \\
                \hline
                Critic Layer Width & \multicolumn{3}{c|}{1024} \\
                \hline
                Dropout Rate & \multicolumn{2}{c|}{0.01} & 0.0 \\
                \hline
                Layer Normalization & \multicolumn{2}{c|}{Yes} & No \\
                \hline
                Batch Renormalization & \multicolumn{2}{c|}{No} & Yes \\
                \hline
                Gradient Steps & \multicolumn{2}{c|}{2} & 1 \\
                \hline
                Policy Delay & \multicolumn{2}{c|}{2} & 3 \\
                \hline
                Target Update Rate & \multicolumn{2}{c|}{0.005} & - \\
                \hline
            \end{tabular}
        \end{table}

    \bibliography{references}